\documentclass[10pt,twocolumn,showpacs,preprintnumbers,amsmath,amssymb,aps,prl,superscriptaddress,longbibliography]{revtex4-2}
\usepackage{appendix}
\usepackage{bbm}
\usepackage{mathrsfs} 
\usepackage{graphicx}
\usepackage{dcolumn}
\usepackage{bm}
\usepackage{braket}
\usepackage{amsmath}
\usepackage{amsfonts}
\usepackage{CJKutf8}
\usepackage[dvipsnames]{xcolor}
\usepackage[colorlinks=true,linkcolor=Blue,urlcolor=BlueViolet,citecolor=BlueViolet]{hyperref}
\usepackage{natbib}
\usepackage{multirow}

\begin{document}

\title{Enhancing Neural-Network Variational Monte Carlo through Basis Transformation}

\author{Zhixuan Liu}
\thanks{These two authors contributed equally to this work.}
\affiliation{State Key Laboratory of Surface Physics and Department of Physics, Fudan University, Shanghai 200433, China}
\affiliation{Shanghai Research Center for Quantum Sciences, Shanghai 201315, China}
\author{Dongheng Qian}
\thanks{These two authors contributed equally to this work.}
\affiliation{State Key Laboratory of Surface Physics and Department of Physics, Fudan University, Shanghai 200433, China}
\affiliation{Shanghai Research Center for Quantum Sciences, Shanghai 201315, China}
\author{Jing Wang}
\thanks{wjingphys@fudan.edu.cn}
\affiliation{State Key Laboratory of Surface Physics and Department of Physics, Fudan University, Shanghai 200433, China}
\affiliation{Shanghai Research Center for Quantum Sciences, Shanghai 201315, China}
\affiliation{Institute for Nanoelectronic Devices and Quantum Computing, Fudan University, Shanghai 200433, China}
\affiliation{Hefei National Laboratory, Hefei 230088, China}

\begin{abstract}
Neural-network variational Monte Carlo (NNVMC) has emerged as a powerful tool for solving quantum many-body problems, yet systematic pathways for improving its accuracy remain largely heuristic. Here, we introduce a physically motivated basis transformation for NNVMC that enhances variational expressivity without increasing the complexity of the neural-network ansatz itself. By formulating the many-body wave function in a Gaussian basis, we introduce a single learnable locality parameter, $\alpha$, that reshapes the target ground state into a more learnable representation. This approach introduces minimal computational overhead and can be readily combined with existing neural-network architectures. Using the three-dimensional homogeneous electron gas as a benchmark, we show that the optimized basis transformation consistently lowers the variational energy for both FermiNet and message-passing neural-network architectures. Notably, for the latter, it enables a more precise determination of the Fermi liquid to Wigner crystal phase transition. More broadly, our results highlight basis transformation as a new route to improving NNVMC in continuous space, showing that accuracy can be enhanced not only by refining the ansatz but also by making the target ground state easier to represent.
\end{abstract}

\maketitle

Solving the quantum many-body problem remains a central challenge in condensed matter physics. Determining ground states, identifying emergent quantum phases, and accurately mapping phase diagrams are essential to understanding material properties. However, the Hilbert space grows exponentially with particle numbers, rendering exact diagonalization intractable and motivating the development of diverse numerical approaches with complementary strengths and limitations. For example, density functional theory is widely used for electronic-structure calculations~\cite{hohenberg1964,kohn1965,jones1989},  but its accuracy is largely limited to weakly correlated regimes. Tensor network methods provide systematically controllable accuracy for low-dimensional systems with area-law entanglement, yet their computational cost grows significantly in higher dimensions~\cite{white1992, schollwock2005, schollwock2011, orus2014, cirac2021}. Quantum Monte Carlo methods can access large systems with high accuracy~\cite{hammond1994,foulkes2001}, yet the notorious fermion sign problem restricts its generic applicability~\cite{loh1990,li2019}. Among these approaches, variational Monte Carlo (VMC) naturally avoids the sign problem~\cite{mcmillan1965}, its accuracy determined by the expressiveness of the trial wavefunction (or ansatz) and the efficiency of the optimization scheme~\cite{toulouse2007}.

Recently, neural quantum states (NQS)—an expressive class of variational ansatz—have revolutionized VMC~\cite{carleo2017, deng2017, nomura2017, sun2022}. Leveraging the universal approximation capability of neural networks and efficient gradient-based optimization via automatic differentiation, NQS have been successfully applied to a broad range of quantum systems~\cite{choo2018, ferrari2019, luo2019, choo2020, hibat2020, robledo2022, pescia2022, li2022ab, viteritti2023, fore2023, hermann2023, wilson2023, lin2023, lange2024, chen2024, li2024, sprague2024, rende2024, zhang2024, qian2025, roth2025, gu2025, valenti2025, sobral2025, zhang2025, chen2025, gerard2025, zaklama2026}. In particular, for continuous-space fermionic systems, architectures such as FermiNet~\cite{pfau2020,spencer2020,cassella2023}, PauliNet~\cite{hermann2020}, and Psiformer~\cite{von2022,teng2025,geier2025,li2025,geier2025b} have achieved state-of-the-art accuracy, establishing neural-network VMC (NNVMC) as a competitive modern approach for \textit{ab initio} electronic-structure calculations.

Despite these advances, a central challenge remains: how can the accuracy of NNVMC be improved efficiently and systematically? A straightforward strategy is to increase the number of variational parameters, thereby enlarging the variational manifold. In practice, however, this brute-force approach often leads to substantially higher computational cost and more difficult optimization, while the resulting accuracy may saturate or even degrade due to overfitting~\cite{dash2025,moss2025}. More fundamentally, blindly increasing the number of parameters lacks clear physical interpretation. This contrasts sharply with tensor network methods, where increasing the bond dimension directly corresponds to accommodating greater entanglement~\cite{cirac2021}. These considerations highlight the need for improvement strategies that are not only efficient and effective but also physically motivated.

In this Letter, we propose enhancing NNVMC through a physically motivated basis transformation. Instead of increasing the complexity of the trial wavefunction itself, we transform the basis in which the Hamiltonian eigenvalue problem is represented, thereby reshaping the ground-state wavefunction that the neural network must approximate. This offers a perspective distinct from conventional ansatz-level improvements. While recent works have explored transformations in discrete space~\cite{moreno2023, cortes2025, kovzic2025, moss2025b}, here we focus on continuous-space fermionic systems and employ a nonorthogonal Gaussian basis characterized by a single parameter $\alpha$ that controls spatial locality. Because only one additional parameter is introduced, the optimization remains stable and the computational overhead is minimal. The method is also architecture-agnostic and can be readily combined with existing NQS ansatz. Using the three-dimensional homogeneous electron gas (3DHEG) as a benchmark system, we show that incorporating $\alpha$ significantly lowers the variational energy for both FermiNet~\cite{cassella2023} and message-passing neural-network (MPNN) architectures~\cite{pescia2024,smith2024}, enabling a more precise determination of the Fermi-liquid (FL) to Wigner crystal (WC) phase transition~\cite{wigner1934,drummond2004,giuliani2008,azadi2023}.

\emph{Basis transformation}---We consider a VMC framework augmented by a basis transformation. The key idea is to introduce a complete, not necessarily orthonormal basis whose parameters are optimized alongside those of the wave function, thereby increasing the variational flexibility and enabling a more accurate approximation to the ground state. Specifically, we define the many-body wave function in real space as
\begin{align}
\tilde{\psi}_{\theta}(\mathbf{r}) = \int d\mathbf{x} \, \psi_{\theta_1}(\mathbf{x}) \, G_{\theta_2}(\mathbf{x},\mathbf{r}),
\label{eq:psi_ansatz}
\end{align}
where $\psi_{\theta_1}(\mathbf{x})$ is a wave function 
defined in an auxiliary coordinate space $\mathbf{x}$, and $G_{\theta_2}(\mathbf{x},\mathbf{r})$ is a kernel that maps the auxiliary coordinates to the physical coordinates $\mathbf{r}$. 
The variational parameters are $\theta\equiv(\theta_1,\theta_2)$, where $\theta_1$ denotes the usual NQS parameters and $\theta_2$ parametrizes the kernel. For fermionic systems, antisymmetry is enforced by requiring $\psi_{\theta_1}(\mathbf{x})$ to be antisymmetric under particle exchange and imposing $G_{\theta_2}(P\mathbf{x},P\mathbf{r})=G_{\theta_2}(\mathbf{x},\mathbf{r})$ for any permutation $P$.

In this work, we employ a Gaussian kernel
\begin{align}
G_{\alpha}(\mathbf{x},\mathbf{r}) = \left(\frac{\alpha}{\pi}\right)^{3n/2} \exp\left(-\alpha \sum_{i=1}^{n} |\mathbf{r}_i - \mathbf{x}_i|^2 \right),
\label{eq:gaussian_kernel}
\end{align}
where $\theta_2 = \{\alpha\}$ is a single parameter and $n$ is the number of electrons. Compared with conventional NNVMC, this introduces only a single additional variational parameter. The parameter $\alpha$ controls the spatial locality of the basis: large $\alpha$ corresponds to a localized basis, and in the limit $\alpha\to\infty$ the kernel approaches a Dirac delta function, recovering the standard real-space basis. The kernel effectively convolves the wave function, and in reciprocal space the convolution corresponds to
\begin{align}
\tilde{\psi}_{\theta}(\mathbf{k}) = \psi_{\theta_1}(\mathbf{k}) \, e^{-\frac{1}{4\alpha}|\mathbf{k}|^2}.
\label{eq:k effect}
\end{align}
which acts as a low-pass filter that suppresses the high-frequency components of the wave function, reflecting the smoothing effect in real space.

Within the VMC framework, $(\theta_1,\alpha)$ are optimized by minimizing the total energy. The basis transformation modifies the Hamiltonian matrix elements to $H_{\alpha}(\mathbf{x},\mathbf{x}') = \langle G_{\alpha}(\mathbf{x})|\hat{H}|G_{\alpha}(\mathbf{x}')\rangle$, while the nonorthogonal basis introduces an overlap matrix $I_\alpha(\mathbf{x},\mathbf{x}') = \langle G_\alpha(\mathbf{x})|G_\alpha(\mathbf{x}')\rangle$. For the Gaussian kernel, the overlap becomes
\begin{align}
I_{\alpha}(\mathbf{x},\mathbf{x}') = \left(\frac{\alpha}{2\pi}\right)^{3n/2} \exp\left(-\frac{\alpha}{2}\left|\mathbf{x} - \mathbf{x}'\right|^2\right),
\label{eq:overlap}
\end{align}
which is strictly positive and defines a normalized Gaussian distribution centered at $\mathbf{x}$ with variance $\sigma=1/\sqrt{\alpha}$. The total energy reads
\begin{equation}
E_\theta = \frac{\int d\mathbf{x} d\mathbf{x}' \, \psi_{\theta_1}^*(\mathbf{x}) \psi_{\theta_1}(\mathbf{x}') H_{\alpha}(\mathbf{x},\mathbf{x}')}
{\int d\mathbf{x} d\mathbf{x}' \, \psi_{\theta_1}^*(\mathbf{x}) \psi_{\theta_1}(\mathbf{x}') I_{\alpha}(\mathbf{x},\mathbf{x}')}.
\label{eq:total_energy}
\end{equation}
The integrals are high dimensional and must be evaluated stochastically. Because the basis is nonorthogonal, the denominator integrand $\int d\mathbf{x}' \psi_{\theta_1}(\mathbf{x}) I_{\alpha}(\mathbf{x},\mathbf{x}')\psi_{\theta_1}(\mathbf{x}')$ is not always positive and direct Monte Carlo sampling is not possible. We therefore construct a positive sampling distribution
\begin{equation}
p_{\theta}(\mathbf{x}) \propto \int d\mathbf{x}' \,\left|\psi^*_{\theta_1}(\mathbf{x})\right| \, \left|\psi_{\theta_1}(\mathbf{x}')\right| \, I_\alpha(\mathbf{x},\mathbf{x}'),
\end{equation}
which exploits the positivity of the Gaussian overlap. With this distribution, the energy can be written as
\begin{equation}
E_\theta = \frac{\mathbb{E}_{p_{\theta}} \big[ S_{\text{L}}(\mathbf{x}) \, E_{\text{L}}(\mathbf{x}) \big]}
{\mathbb{E}_{p_{\theta}} \big[ S_{\text{L}}(\mathbf{x}) \big]},
\label{eq:energy_monte_carlo}
\end{equation}
where $E_{\text{L}}(\mathbf{x})$ is the generalized local energy and $S_{\text{L}}(\mathbf{x})$ is the local sign 
\begin{eqnarray}
E_{\text{L}}(\mathbf{x}) &=& \frac{\int d\mathbf{x}' \, H_{\alpha}(\mathbf{x},\mathbf{x}') \psi_{\theta_1}(\mathbf{x}')}{\int d\mathbf{x}' \, I_{\alpha}(\mathbf{x},\mathbf{x}') \psi_{\theta_1}(\mathbf{x}')},
\nonumber
\\
S_{\text{L}}(\mathbf{x}) &=& \frac{\int d\mathbf{x}' \, I_{\alpha}(\mathbf{x},\mathbf{x}') \,\left|\psi_{\theta_1}(\mathbf{x}')\right| \, \operatorname{sgn}\big[\psi^*_{\theta_1}(\mathbf{x})\psi_{\theta_1}(\mathbf{x}')\big]}
{\int d\mathbf{x}' \, I_{\alpha}(\mathbf{x},\mathbf{x}')\left|\psi_{\theta_1}(\mathbf{x}')\right|}.
\nonumber
\end{eqnarray}
Here, $\operatorname{sgn}(C)$ denotes the phase of complex number $C$.

The physical meaning of these quantities is clear. The ground state satisfies the generalized eigenvalue problem with a nonorthogonal basis $\int d\mathbf{x}' H(\mathbf{x},\mathbf{x}')\psi_{\mathrm{GS}}(\mathbf{x}') = E_{\mathrm{GS}} \int d\mathbf{x}' I(\mathbf{x},\mathbf{x}')\psi_{\mathrm{GS}}(\mathbf{x}')$. $E_{\text{L}}(\mathbf{x})$ therefore becomes spatially constant and equals $E_{\mathrm{GS}}$ for the exact ground state, a property that stabilizes parameter optimization. $S_{\text{L}}(\mathbf{x})$ captures the phase information of the wave function introduced by absolute value sampling, which has no counterpart in standard VMC. Interpreting $I_{\alpha}(\mathbf{x},\mathbf{x}') |\psi_{\theta_1}(\mathbf{x}')|$ as a probability distribution, $S_{\text{L}}(\mathbf{x})$ represents an average sign difference between the wave function at $\mathbf{x}'$ relative to $\mathbf{x}$. Notably, since $I_{\alpha}(\mathbf{x},\mathbf{x}')$ is Gaussian, these inner integrals over $\mathbf{x}'$ can be efficiently evaluated by sampling $\mathbf{x}' \sim \mathcal{N}(\mathbf{x}, 1/\sqrt{\alpha})$.

\emph{Optimization strategy---}The parameters $(\theta_1,\alpha)$ are optimized by minimizing \(E_\theta\). Differentiating Eq.~\eqref{eq:energy_monte_carlo} yields
\begin{align}
\nabla_{\theta} E_\theta = 2\,\mathrm{Re} \left\{ \frac{\mathbb{E}_{p_{\theta}(\mathbf{x})} \big[ O^*_{\theta}(\mathbf{x})\, \epsilon(\mathbf{x}) \, S_{\text{L}}(\mathbf{x}) \big]}
{\mathbb{E}_{p_{\theta}(\mathbf{x})} \big[S_{\text{L}}(\mathbf{x})\big]}\right\},
\label{eq:grad_theta1}
\end{align}
where $\epsilon(\mathbf{x}) = E_{\text{L}}(\mathbf{x}) - E_\theta$. For the wave function parameters, $O_{\theta_1}(\mathbf{x})\equiv \nabla_{\theta_1}\psi_{\theta_1}(\mathbf{x})/\psi_{\theta_1}(\mathbf{x})$, while for the basis parameter, $O_{\alpha}(\mathbf{x}) = -(1/4\alpha^2) \nabla_{\mathbf{x}}^2 \psi(\mathbf{x})/\psi(\mathbf{x})$. We have used the identity $\nabla_{\alpha}\tilde{\psi}_{\theta}(\mathbf{r}) = 
-(1/4\alpha^2)\int d\mathbf{x} \, G_{\alpha}(\mathbf{x},\mathbf{r}) \nabla_{\mathbf{x}}^2 \psi(\mathbf{x})$,
which implies that an infinitesimal change \(\alpha \to \alpha + \delta\alpha\) acts as \(\psi(\mathbf{x}) \to \psi(\mathbf{x}) - (\delta\alpha/4\alpha^2)\nabla_{\mathbf{x}}^2 \psi(\mathbf{x})\). In practice, we employ stochastic reconfiguration to accelerate optimization~\cite{sorella2001,chen2024,goldshlager2024}. Details are provided in the Supplemental Material~\cite{supp}.

A naive simultaneous optimization of $\theta_1$ and $\alpha$ is prone to numerical instability. This stems from a fundamental coupling between the basis locality and the statistical variance of the gradient estimates. Specifically, the Monte Carlo evaluation of Eq.~(\ref{eq:grad_theta1}) requires sampling $\mathbf{x}'$ from the Gaussian distribution $\mathcal{N}(\mathbf{x}, 1/\sqrt{\alpha})$. When \(\alpha\) is small, the sampling distribution becomes highly nonlocal, leading to large statistical errors in the gradient estimates. If \(\theta_1\) and \(\alpha\) are updated concurrently, the poor fit of the initial ansatz $\psi_{\theta_1}(\mathbf{x})$ often causes $\alpha$ to decrease prematurely. This ``delocalization'' of the basis increases gradient noise, which in turn prevents $\theta_1$ from converging toward the ground state, causing $\alpha$ to shrink even more—a vicious cycle that often leads to optimization failure.

\begin{figure}[t]
\centering
\includegraphics[width=0.8\linewidth]{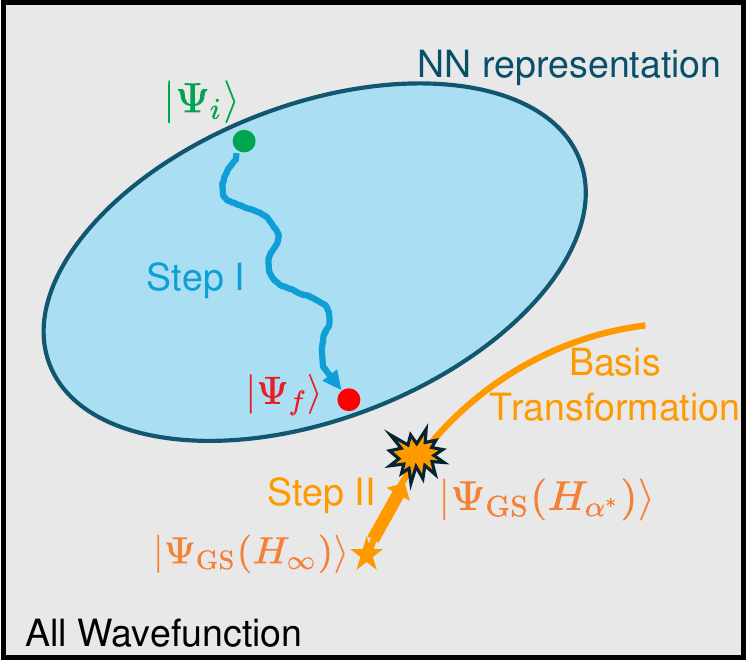}
\caption{Schematic of the two-step optimization. In Step I: conventional VMC is performed with a fixed basis ($\alpha \to \infty$), optimizing the wave function from the initial state $|\Psi_i\rangle$ to the converged final state $|\Psi_f\rangle$. In Step II, the wave function parameters are fixed and the basis parameter $\alpha$ is optimized toward the optimal value $\alpha^*$. This effectively shifts the Hamiltonian and its ground state from $|\Psi_{\mathrm{GS}}(H_{\alpha=\infty})\rangle$ to $|\Psi_{\mathrm{GS}}(H_{\alpha^{*}})\rangle$, reducing the distance between the optimized wave function $|\Psi_f\rangle$ and the target ground state $|\Psi_{\mathrm{GS}}(H_{\alpha^{*}})\rangle$.}
\label{fig:framework}
\end{figure}

To address this issue, we adopt a two-step optimization framework illustrated in Fig.~\ref{fig:framework}, 
\begin{itemize}
    \item Step I: wavefunction pre-training. $\alpha$ is fixed to a large value ($\alpha\to\infty$) and only $\theta_1$ is optimized, effectively reducing the method to conventional VMC in a local real-space basis and allowing $\theta_1$ to be optimized stably. This drives the initial state \(|\Psi_{i}\rangle\) toward the exact ground state \(|\Psi_{\mathrm{GS}}(H_{\alpha=\infty})\rangle\), yielding \(|\Psi_{f}\rangle\) and the corresponding energy \(E^\prime\). 
    \item Step II: basis refinement. With well-trained $\theta_1$ fixed, we enable the update of $\alpha$. The pre-trained wavefunction prevents $\alpha$ from collapsing to excessively small values during subsequent learning.
\end{itemize}
This second step is the key difference from conventional VMC: varying $\alpha$ effectively modifies the Hamiltonian in the auxiliary representation, shifting the target ground state from $|\Psi_{\mathrm{GS}}(H_{\alpha=\infty})\rangle$ to $|\Psi_{\mathrm{GS}}(H_{\alpha^*})\rangle$. This reduces the distance between the trained wave function $|\Psi_{f}\rangle$ and the optimal ground state $|\Psi_{\mathrm{GS}}(H_{\alpha^*})\rangle$ in function space, resulting in a lower total energy $E$. The energy difference $E - E^\prime$ quantifies the improvement due to the basis transformation.

\emph{Benchmarking 3DHEG}---To demonstrate the effectiveness of our method, we study 3DHEG, a paradigmatic model of interacting fermions. In Hartree atomic units its Hamiltonian is~\cite{mahan2013}
\begin{equation}
H = -\frac{1}{2} \sum_i \nabla_i^2 + \sum_{i<j} \frac{1}{|\mathbf{r}_i - \mathbf{r}_j|} + \text{b.g.},
\label{eq:heg_hamiltonian}
\end{equation}
where b.g. denotes the neutralizing background contribution~\cite{Kwon1988}. The system is characterized by a single dimensionless parameter—the Wigner–Seitz radius 
$r_s$ (in units of the Bohr radius $a_B$). We employ periodic boundary conditions and evaluate Coulomb interactions using the Ewald summation technique~\cite{Ewald1921,Fraser1996,Abdulnour1996}. Within the Gaussian basis transformation, the Hamiltonian matrix elements $H_\alpha(\mathbf{x},\mathbf{x}')$ can be computed analytically; details are given in the Supplemental Material~\cite{supp}.

\begin{figure}[b]
\centering
\includegraphics[width=1.0\linewidth]{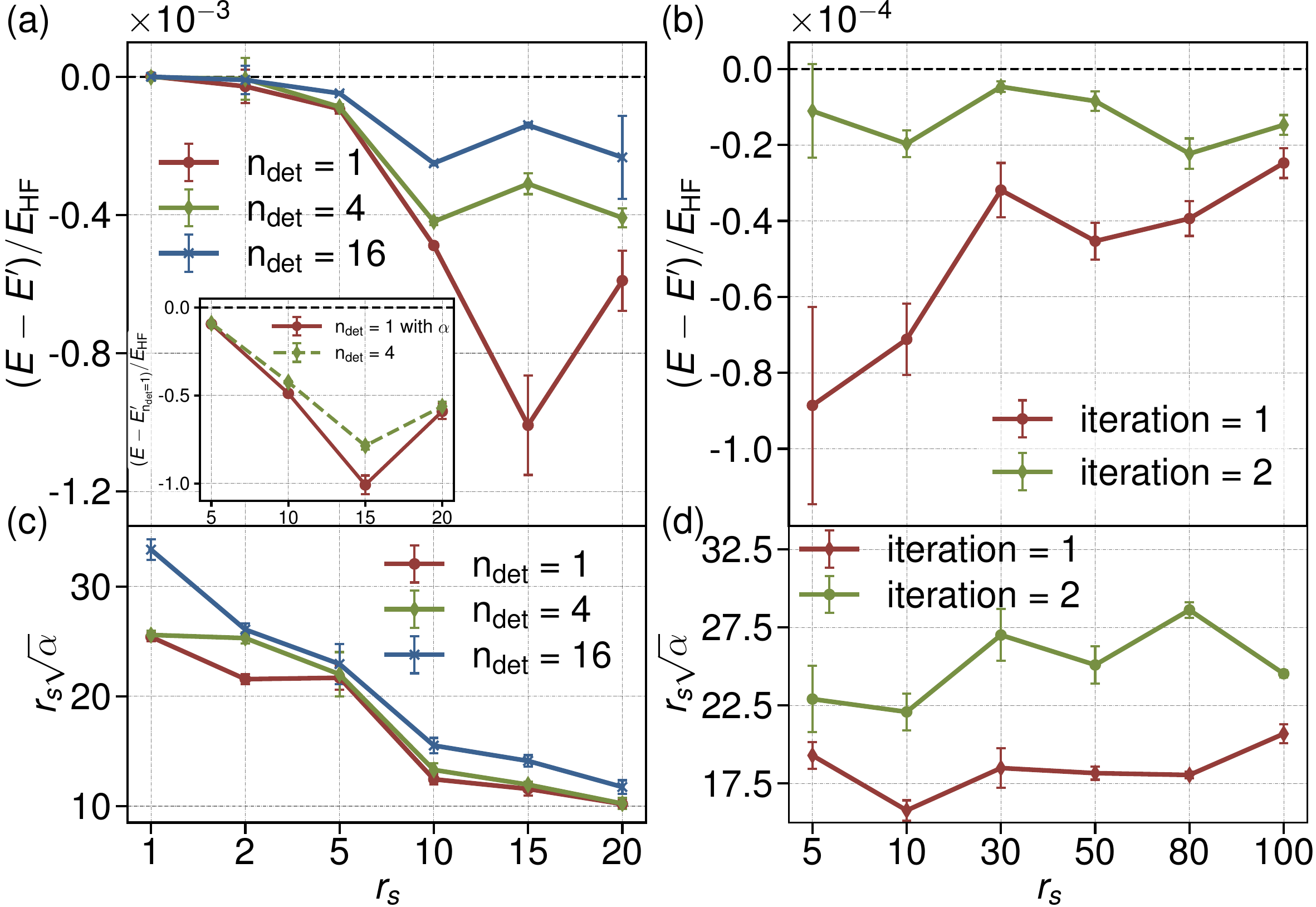}
\caption{Comparison of ground-state energies with and without the Gaussian basis parameter \(\alpha\). (a) Relative energy difference \((E-E^\prime)/E_{\text{HF}}\) for FermiNet with different numbers of Slater determinants \(n_{\text{det}}\), where \(E_{\text{HF}}\) is the energy of the free electron gas wave function. The inset compares the energy gains from introducing \(\alpha\) with that obtained by increasing \(n_{\text{det}}\) from 1 to 4 (without \(\alpha\)). (b) Same quantity as in (a) for the MPNN architecture with different numbers of message-passing iterations. (c),(d) Optimized dimensionless basis parameter 
$r_s\sqrt{\alpha}$ for the corresponding FermiNet (c) and MPNN (d) calculations.}
\label{fig:energy_difference}
\end{figure}

We benchmark two widely used neural-network architectures for the 3DHEG under different system settings: FermiNet for an unpolarized 14-electron system in a cubic cell, and a message-passing neural network (MPNN) for an unpolarized 36-electron system in a rectangular cell with aspect ratio 
$3:3:2$. For the MPNN, we consider two choices of reference states, plane waves (PW) and Gaussian orbitals (GO).

The variational energy differences with and without the basis transformation are shown in Fig.~\ref{fig:energy_difference}(a) for FermiNet and Fig.~\ref{fig:energy_difference}(b) for MPNN using the PW reference states. Several features are observed. First, the basis transformation consistently reduces the variational energy for both architectures across the entire range of $r_s$, demonstrating that the approach is architecture-agnostic and universal. Second, introducing the single parameter $\alpha$ provides a highly efficient way to increase the flexibility of the ansatz. As shown in the inset of Fig.~\ref{fig:energy_difference}(a), adding $\alpha$ yields a larger energy gain at $r_s = 5, 10, 20$ than increasing the number of Slater determinants in FermiNet from $n_{\text{det}}=1$ to $4$, which introduces more than $10^4$ additional parameters.

\begin{figure}[t]
\centering
\includegraphics[width=0.95\linewidth]{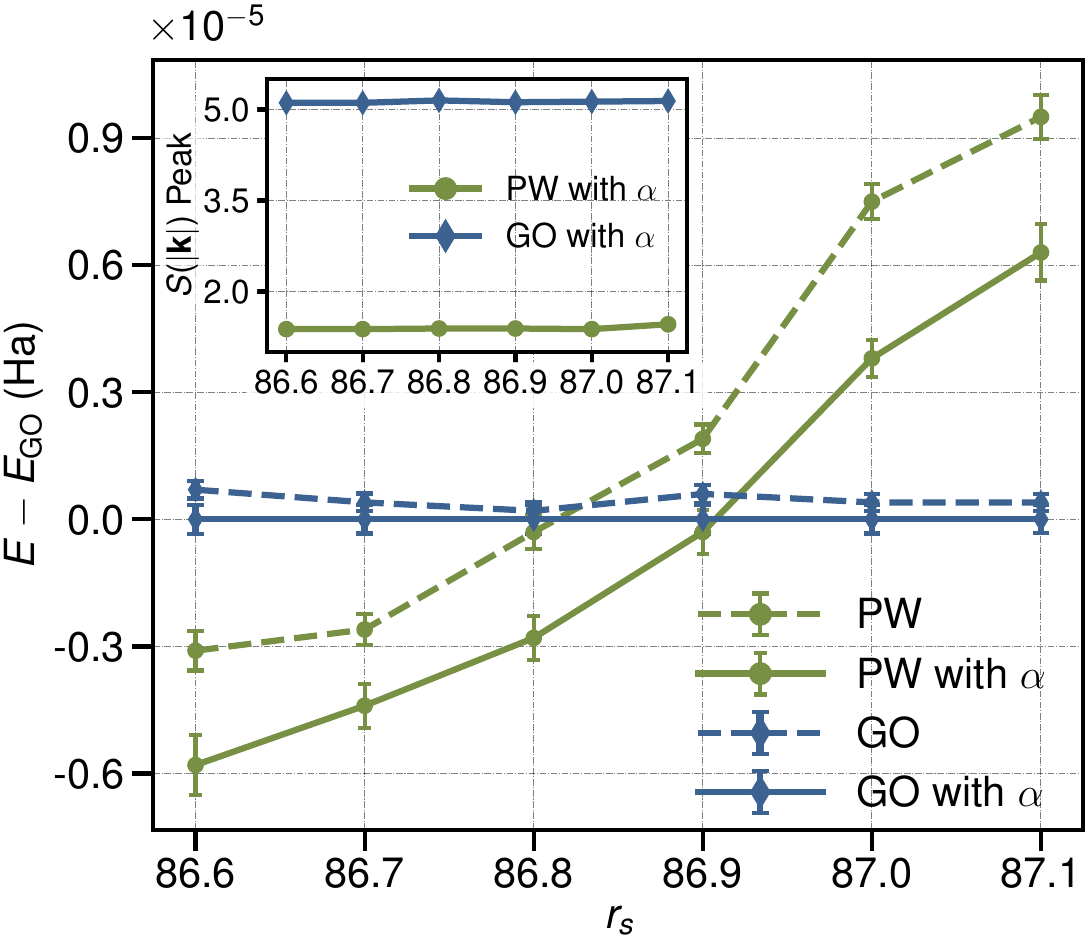}
\caption{Ground-state energy as a function of $r_s$ for the MPNN architecture in a $36$-electron system with plane-wave (PW) and Gaussian-orbital (GO) reference states, with and without the Gaussian basis parameter $\alpha$. The inset shows the peak value of the radially averaged static structure factor $S(|\mathbf{k}|)$. A large value indicates a pronounced Bragg peak and is the signature of the Wigner crystal (WC) phase, confirming that the PW reference state remains in the Fermi liquid (FL) phase while the GO reference state favors the WC phase.}
\label{fig:phase_change}
\end{figure}

The optimized values of the dimensionless parameter 
$r_s\sqrt{\alpha}$ are shown in Figs.~\ref{fig:energy_difference}(c) and~\ref{fig:energy_difference}(d). A clear trend emerges: smaller values of \(r_s\sqrt{\alpha}\) correlate with larger energy improvements, consistent with the expectation that smaller $\alpha$ corresponds to a more nonlocal basis and thus a stronger modification of the original wave function. As the network complexity increases, the optimal $r_s\sqrt{\alpha}$ shifts to larger values, indicating that a more accurate baseline ansatz requires less correction from the basis transformation. The dependence on $r_s$ differs qualitatively between two architectures: for FermiNet the optimal $r_s\sqrt{\alpha}$ increases with $r_s$, suggesting that the baseline ansatz becomes less accurate at low density and therefore benefits more from the transformation, whereas for the MPNN the optimal value of $r_s\sqrt{\alpha}$ remains nearly constant across the range of $r_s$, implying a more uniform baseline performance as the density varies.

The improved energies also enable a more precise determination of the FL-WC transition point. We focus on the MPNN architecture and employ both PW and GO reference states. As shown in Fig.~\ref{fig:phase_change}, over the \(r_s\) range considered, the PW reference state consistently yields the FL phase, while the GO reference state stabilizes the WC phase, as confirmed by the Bragg peak in the static structure factor $S(|\mathbf{k}|)$ shown in the inset. The crossing of the corresponding energies therefore determines the transition point. Incorporating the basis transformation lowers the energy more strongly for the PW reference state than for the GO reference state, shifting the estimated FL–WC transition to larger $r_s$ by $|\delta r_s| \approx 0.1$.

\begin{figure}[b]
\centering
\includegraphics[width=1.0\linewidth]{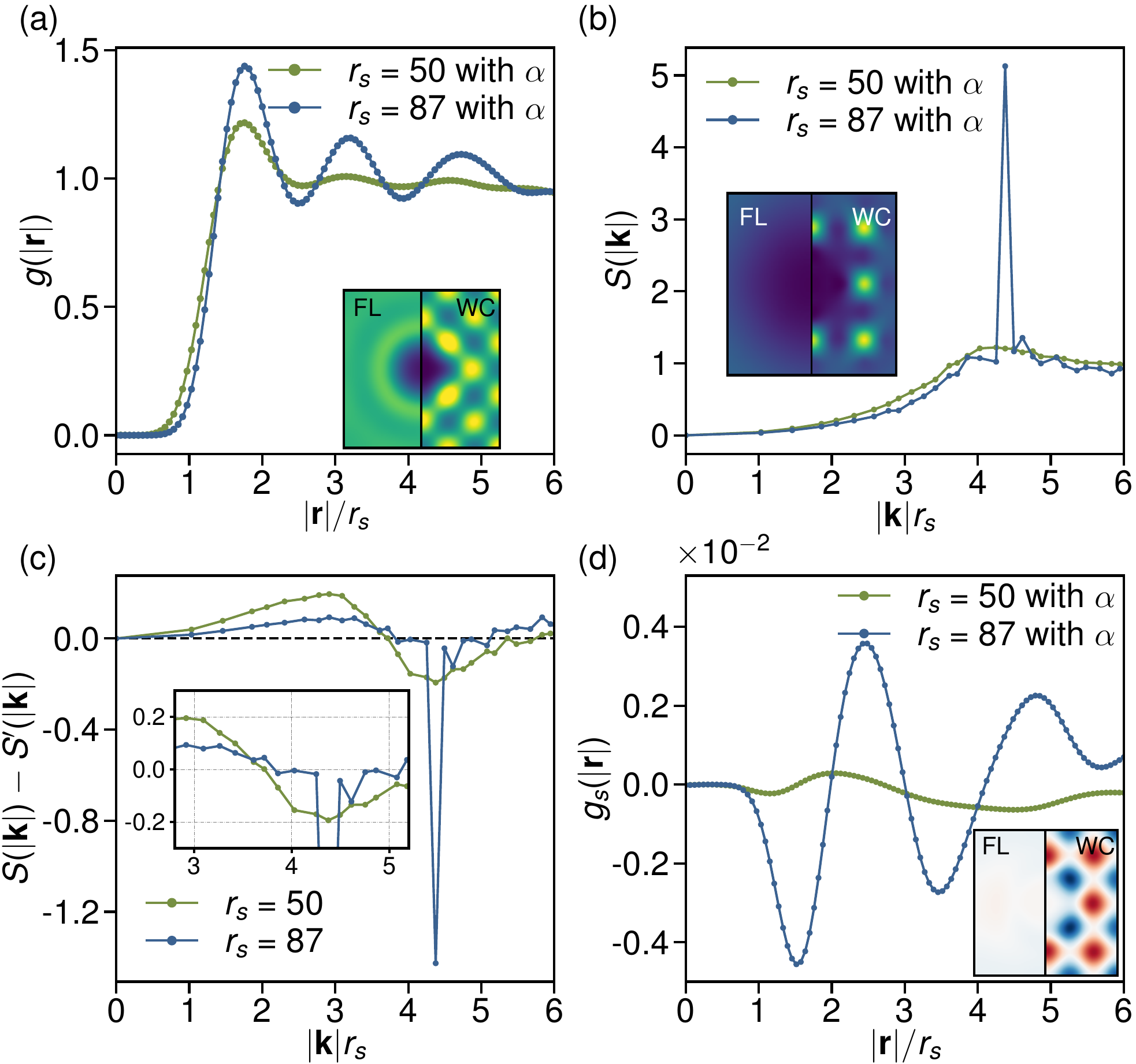}
\caption{Electron-electron correlations after basis transformation for $r_s=50$ and $r_s=87$. (a) Pair correlation function $g(|\mathbf{r}|)$. (b) Spin-averaged static structure factor $S(|\mathbf{k}|)$. (c) Difference between the structure factor with basis transformation $S(|\mathbf{k}|)$ and that without transformation $S^{\prime}(|\mathbf{k}|)$. (d) Spin-spin pair correlation function $g_s(|\mathbf{r}|)$.  
Main panels show radially averaged quantities. Insets in (a), (b), and (d) display two-dimensional projections; the left halves correspond to the Fermi-liquid (FL) phase ($r_s=50$, PW reference) and the right halves to the WC phase ($r_s=87$, GO reference). Insets in (a) and (d) are summed over the $z=0$ and $z=a_{\text{BCC}}/2$ planes, where $a_{\text{BCC}}$ is the conventional cell lattice constant of the BCC lattice; while the inset in (b) is summed over the $k_z=0$ and $k_z=k_{\text{BCC}}/2$ planes in reciprocal space, where $k_{\text{BCC}}$ is the corresponding conventional reciprocal lattice constant.}
\label{fig:correlation_details}
\end{figure}

Finally, ground state observables can also be evaluated efficiently within the transformed basis~\cite{supp}.Figures~\ref{fig:correlation_details}(a) and~\ref{fig:correlation_details}(b) show the pair correlation function $g(|\mathbf{r}|)$ and the static structure factor $(S(|\mathbf{k}|)$ for representative densities $r_s = 50$ (PW reference state) and $r_s = 87$ (GO reference state). At $r_s = 87$, $g(|\mathbf{r}|)$ exhibits long-range oscillations and $S(|\mathbf{k}|)$ displays sharp Bragg peaks, consistent with a body-centered cubic WC. In contrast, the state at
$r_s = 50$ shows a smooth $S(|\mathbf{k}|)$ characteristic of a FL. We further examine the effect of the Gaussian basis transformation on \(S(|\mathbf{k}|)\), the transformation enhances \(S(|\mathbf{k}|)\) small momenta and suppresses it at large momenta [Fig.~\ref{fig:correlation_details}(c)], 
consistent with the low-pass filtering effect implied by
Eq.~(\ref{eq:k effect}). We also examine spin correlations. As shown in Fig.~\ref{fig:correlation_details}(d), the WC phase at 
$r_s = 87$ exhibits pronounced long-range oscillations in the spin-spin correlation function, reflecting antiferromagnetic ordering in the crystalline phase.

\emph{Discussions}---We have introduced a nonorthogonal basis transformation as a physically motivated route to improving NNVMC, and demonstrated its effectiveness in the 3DHEG case. Rather than increasing the complexity of the NQS ansatz itself, we show that accuracy can be systematically improved by making the target ground state ``easier'' for the network to represent. This perspective opens up a complementary route for improving neural-network approaches to quantum many-body problems.

The framework is general and can be readily applied to other quantum many-body systems. In systems where competing phases are separated by very small energy differences, such as those exhibiting superconductivity~\cite{roth2025}, we expect that the energy gain enabled by the basis transformation may lead to more substantial modifications of the phase diagram, rather than merely shifting the phase boundary as in the 3DHEG studied here. Another particularly interesting direction concerns problems with nonlocal potentials, such as those involving nonlocal pseudopotentials~\cite{li2022a, fu2025}. In such cases, the evaluation of the local energy already intrinsically requires integration, making our approach more natural to implement.

More broadly, beyond using basis transformation as an efficient way to enhance expressivity, it is also worthwhile to explore whether it can improve the optimization landscape itself, potentially making the true ground state more accessible during training. Exploring this possibility would generally require the simultaneous optimization of both the wavefunction and basis parameters, which in turn hinges on a more efficient evaluation of the local energy. We leave these directions for future work. We hope that the present work will stimulate further investigation of basis-engineering strategies for neural-network approaches to quantum many-body problems.

\begin{acknowledgments}
\emph{Acknowledgments}---This work is supported by the Natural Science Foundation of China through Grant No.~12350404, the Quantum Science and Technology-National Science and Technology Major Project through Grant No.~2021ZD0302600, the Science and Technology Commission of Shanghai Municipality under Grants No.~23JC1400600, No.~24LZ1400100 and No.~2019SHZDZX01, and it is sponsored by the ``Shuguang Program'' supported by the Shanghai Education Development Foundation and Shanghai Municipal Education Commission.
\end{acknowledgments}

\end{document}


\title{Supplemental Material for ``Enhancing Neural-Network Variational Monte Carlo through Basis Transformation''}

\author{Zhixuan Liu}
\thanks{These two authors contributed equally to this work.}
\affiliation{State Key Laboratory of Surface Physics and Department of Physics, Fudan University, Shanghai 200433, China}
\affiliation{Shanghai Research Center for Quantum Sciences, Shanghai 201315, China}
\author{Dongheng Qian}
\thanks{These two authors contributed equally to this work.}
\affiliation{State Key Laboratory of Surface Physics and Department of Physics, Fudan University, Shanghai 200433, China}
\affiliation{Shanghai Research Center for Quantum Sciences, Shanghai 201315, China}
\author{Jing Wang}
\thanks{Contact author: wjingphys@fudan.edu.cn}
\affiliation{State Key Laboratory of Surface Physics and Department of Physics, Fudan University, Shanghai 200433, China}
\affiliation{Shanghai Research Center for Quantum Sciences, Shanghai 201315, China}
\affiliation{Institute for Nanoelectronic Devices and Quantum Computing, Fudan University, Shanghai 200433, China}
\affiliation{Hefei National Laboratory, Hefei 230088, China}

\maketitle

\tableofcontents

\section{Basis transformation in VMC}

In this section, we provide more analytical details regarding the evaluation of observables and energy gradients, the acceleration of parameter updates, and the analytical forms of various operator matrix elements when the basis transformation is applied.

\subsection{Wave function in Gaussian basis}

We first present the explicit form of the wave function after introducing the basis transformation. The transformed wave function is written as
\[
\tilde{\psi}_{\theta}(\mathbf{r}) =
\int d\mathbf{x} \, \psi_{\theta_1}(\mathbf{x}) \,
G_{\theta_2}(\mathbf{x},\mathbf{r}) ,
\]
where \(\psi_{\theta_1}(\mathbf{x})\) is a wave function defined in an auxiliary coordinate space \(\mathbf{x}\), parametrized by a neural network with parameters \(\theta_1\). The kernel \(G_{\theta_2}(\mathbf{x},\mathbf{r})\) maps the auxiliary coordinates \(\mathbf{x}\) to the physical coordinates \(\mathbf{r}\), and is parametrized by \(\theta_2\).

In this work, we choose a Gaussian basis for the kernel,
\[
G_{\alpha}(\mathbf{x},\mathbf{r}) =
\left(\frac{\alpha}{\pi}\right)^{3n/2}
\exp\left[
-\alpha \sum_{i=1}^{n} |\mathbf{r}_i-\mathbf{x}_i|^2
\right],
\]
where \(n\) is the number of electrons and \(\alpha\) is a learnable scalar parameter controlling the width of the Gaussian basis. We therefore denote the full set of variational parameters by
\[
\theta \equiv \{\theta_1,\theta_2\}=\{\theta_1,\alpha\}.
\]
For fermionic systems, antisymmetry under particle exchange is ensured by requiring \(\psi_{\theta_1}(\mathbf{x})\) to be antisymmetric, while the kernel must satisfy
\[
G_{\alpha}(P\mathbf{x},P\mathbf{r}) = G_{\alpha}(\mathbf{x},\mathbf{r})
\]
for any particle permutation \(P\). This invariance condition is automatically satisfied by the Gaussian kernel above.

\subsection{Evaluation of Observables}

With the analytical form of the transformed wave function at hand, we now describe the evaluation of observables. For a physical observable \(\hat{O}\), its expectation value with respect to \(\tilde{\psi}_{\theta}(\mathbf{r})\) is given by
\begin{equation}
\langle \hat{O} \rangle_\theta =
\frac{
\displaystyle
\int d\mathbf{x}\, d\mathbf{x}' \,
\psi_{\theta_1}^*(\mathbf{x})
\psi_{\theta_1}(\mathbf{x}')
O_{\alpha}(\mathbf{x},\mathbf{x}')
}{
\displaystyle
\int d\mathbf{x}\, d\mathbf{x}' \,
\psi_{\theta_1}^*(\mathbf{x})
\psi_{\theta_1}(\mathbf{x}')
I_{\alpha}(\mathbf{x},\mathbf{x}')
},
\label{eq:expectation_value}
\end{equation}
where
$
O_{\alpha}(\mathbf{x},\mathbf{x}')
=
\langle G_{\alpha}(\mathbf{x}) | \hat{O} | G_{\alpha}(\mathbf{x}') \rangle
$
denotes the matrix element of \(\hat{O}\) in the transformed basis, and
$
I_{\alpha}(\mathbf{x},\mathbf{x}')
=
\langle G_{\alpha}(\mathbf{x}) | G_{\alpha}(\mathbf{x}') \rangle
$
is the corresponding overlap matrix element. For the Gaussian basis, the overlap takes the analytical form
\begin{equation}
I_{\alpha}(\mathbf{x},\mathbf{x}')
=
\left(\frac{\alpha}{2\pi}\right)^{3n/2}
\exp\left[
-\frac{\alpha}{2}
|\mathbf{x}-\mathbf{x}'|^2
\right],
\label{eq:overlap_distribution}
\end{equation}
which is strictly positive and defines a normalized Gaussian distribution in \(\mathbf{x}'\), centered at \(\mathbf{x}\), with standard deviation \(1/\sqrt{\alpha}\) for each coordinate.

Equation~\eqref{eq:expectation_value} involves integrals over both \(\mathbf{x}\) and \(\mathbf{x}'\), which we evaluate by Monte Carlo integration. Following the principle of importance sampling, we introduce the probability distribution
\begin{equation}
q_{\theta}(\mathbf{x},\mathbf{x}')
\propto
|\psi_{\theta_1}(\mathbf{x})|^2
I_{\alpha}(\mathbf{x},\mathbf{x}') .
\end{equation}
In the limit \(\alpha \to \infty\), the overlap kernel approaches a delta function, and the distribution \(q_{\theta}(\mathbf{x},\mathbf{x}')\) reduces to
\[
q_{\theta}(\mathbf{x},\mathbf{x}')
\to
|\psi_{\theta_1}(\mathbf{x})|^2
\delta(\mathbf{x}-\mathbf{x}') ,
\]
thereby recovering the standard variational Monte Carlo expression for expectation values.

Substituting this importance-sampling distribution into Eq.~\eqref{eq:expectation_value}, we obtain
\begin{equation}
\langle \hat{O} \rangle_\theta =
\frac{
\mathbb{E}_{q_{\theta}(\mathbf{x},\mathbf{x}')}
\left[
\frac{O_{\alpha}(\mathbf{x},\mathbf{x}')}
     {I_{\alpha}(\mathbf{x},\mathbf{x}')}
\frac{\psi_{\theta_1}(\mathbf{x}')}
     {\psi_{\theta_1}(\mathbf{x})}
\right]
}{
\mathbb{E}_{q_{\theta}(\mathbf{x},\mathbf{x}')}
\left[
\frac{\psi_{\theta_1}(\mathbf{x}')}
     {\psi_{\theta_1}(\mathbf{x})}
\right]
}.
\label{eq:mc_expectation_value}
\end{equation}
This expression allows the numerator and denominator to be estimated separately using Monte Carlo samples drawn from \(q_{\theta}(\mathbf{x},\mathbf{x}')\).

It remains to specify how to sample the pair \((\mathbf{x},\mathbf{x}')\) from \(q_{\theta}(\mathbf{x},\mathbf{x}')\). Since the overlap kernel \(I_{\alpha}(\mathbf{x},\mathbf{x}')\) is normalized with respect to \(\mathbf{x}'\), the distribution can be factorized as
\begin{equation}
q_{\theta}(\mathbf{x},\mathbf{x}')
=
q_{\theta_1}(\mathbf{x})\,
q_{\alpha}(\mathbf{x}'|\mathbf{x}),
\end{equation}
where
\begin{equation}
q_{\theta_1}(\mathbf{x})
=
\frac{
|\psi_{\theta_1}(\mathbf{x})|^2
}{
\int d\mathbf{x}\,
|\psi_{\theta_1}(\mathbf{x})|^2
},
\qquad
q_{\alpha}(\mathbf{x}'|\mathbf{x})
=
I_{\alpha}(\mathbf{x},\mathbf{x}') .
\end{equation}
The conditional distribution \(q_{\alpha}(\mathbf{x}'|\mathbf{x})\) is a Gaussian distribution. Sampling can therefore be performed in two steps. First, we sample \(\mathbf{x}\) from \(q_{\theta_1}(\mathbf{x})\) using Markov chain Monte Carlo (MCMC). Second, conditioned on the sampled \(\mathbf{x}\), we draw \(\mathbf{x}'\) directly from the Gaussian distribution \(q_{\alpha}(\mathbf{x}'|\mathbf{x})\). The resulting pair \((\mathbf{x},\mathbf{x}')\) is then distributed according to \(q_{\theta}(\mathbf{x},\mathbf{x}')\).

\subsection{Evaluation of Energy Gradient}

In the VMC framework, we optimize the variational parameters by gradient descent in order to minimize the energy. To derive the energy gradient, we first consider the parameter derivatives of the transformed wave function. The parameters \(\theta_1\) enter only through the auxiliary-space wave function \(\psi_{\theta_1}(\mathbf{x})\), while the Gaussian kernel \(G_\alpha(\mathbf{x},\mathbf{r})\) is independent of \(\theta_1\). Therefore, differentiating the transformed wave function with respect to \(\theta_1\) gives
\begin{equation}
  \nabla_{\theta_1} \tilde{\psi}_\theta(\mathbf{r})
  =
  \int d\mathbf{x} \,
  \nabla_{\theta_1}\psi_{\theta_1}(\mathbf{x}) \,
  G_{\alpha}(\mathbf{x},\mathbf{r}) .
\end{equation}

The parameter \(\alpha\) affects the transformed wave function in a different way from \(\theta_1\). While \(\theta_1\) directly parametrizes the auxiliary-space wave function \(\psi_{\theta_1}(\mathbf{x})\), the parameter \(\alpha\) controls the width of the Gaussian kernel \(G_\alpha(\mathbf{x},\mathbf{r})\). Nevertheless, its effect can be recast as an effective modification of \(\psi_{\theta_1}(\mathbf{x})\) under the integral. Explicitly, we have
\begin{equation}
\begin{aligned}
\nabla_{\alpha} \tilde{\psi}_\theta(\mathbf{r})
&=
\int d\mathbf{x} \,
\psi_{\theta_1}(\mathbf{x})\,
\nabla_{\alpha}G_{\alpha}(\mathbf{x},\mathbf{r})
\\
&=
\left(\frac{\alpha}{\pi}\right)^{3n/2}
\int d\mathbf{x} \,
\psi_{\theta_1}(\mathbf{x})
\left(
\frac{3n}{2\alpha}
-
|\mathbf{r}-\mathbf{x}|^2
\right)
\exp\left[-\alpha|\mathbf{r}-\mathbf{x}|^2\right]
\\
&=
-\frac{1}{4\alpha^2}
\left(\frac{\alpha}{\pi}\right)^{3n/2}
\int d\mathbf{x} \,
\psi_{\theta_1}(\mathbf{x})\,
\nabla_{\mathbf{x}}^2
\exp\left[-\alpha|\mathbf{r}-\mathbf{x}|^2\right]
\\
&=
-\frac{1}{4\alpha^2}
\int d\mathbf{x} \,
\nabla_{\mathbf{x}}^2\psi_{\theta_1}(\mathbf{x})\,
G_{\alpha}(\mathbf{x},\mathbf{r}) ,
\end{aligned}
\label{eq:alpha_update}
\end{equation}
where integration by parts has been used in the last step, assuming that the boundary terms vanish.

Equation~\eqref{eq:alpha_update} shows that an infinitesimal change \(\alpha \to \alpha+d\alpha\) induces the following effective variation of the auxiliary-space wave function:
\begin{equation}
\psi_{\theta_1}(\mathbf{x})
\to
\psi_{\theta_1}(\mathbf{x})
-
\frac{d\alpha}{4\alpha^2}
\nabla_{\mathbf{x}}^2\psi_{\theta_1}(\mathbf{x}) .
\end{equation}
In this sense, the effects of varying \(\theta_1\) and \(\alpha\) can be expressed in a unified form as variations of the auxiliary-space wave function. Under an infinitesimal parameter update \((d\theta_1,d\alpha)\), the induced variation is
\begin{equation}
\delta\psi(\mathbf{x})
=
\nabla_{\theta_1}\psi_{\theta_1}(\mathbf{x})\cdot d\theta_1
-
\frac{1}{4\alpha^2}
\nabla_{\mathbf{x}}^2\psi_{\theta_1}(\mathbf{x}) \cdot d\alpha.
\label{eq:delta_psi}
\end{equation}


With the response of the transformed wave function to infinitesimal parameter variations established, we now derive the corresponding energy gradient. The variational energy is obtained by taking \(\hat{O}=\hat{H}\) in Eq.~\eqref{eq:expectation_value}, namely
\begin{equation}
E_\theta =
\frac{
\displaystyle
\int d\mathbf{x}\, d\mathbf{x}' \,
\psi_{\theta_1}^*(\mathbf{x})
\psi_{\theta_1}(\mathbf{x}')
H_{\alpha}(\mathbf{x},\mathbf{x}')
}{
\displaystyle
\int d\mathbf{x}\, d\mathbf{x}' \,
\psi_{\theta_1}^*(\mathbf{x})
\psi_{\theta_1}(\mathbf{x}')
I_{\alpha}(\mathbf{x},\mathbf{x}')
},
\label{eq:energy_expectation}
\end{equation}
where
\[
H_{\alpha}(\mathbf{x},\mathbf{x}')
=
\langle G_{\alpha}(\mathbf{x})|\hat{H}|G_{\alpha}(\mathbf{x}')\rangle
\]
is the Hamiltonian matrix element in the transformed basis.

Under an infinitesimal variation \(\delta\psi(\mathbf{x})\) of the auxiliary-space wave function, the corresponding first-order variation of the energy is
\begin{equation}
\delta E_\theta
=
2\,\mathrm{Re}
\left\{
\int d\mathbf{x}\,
\delta\psi^*(\mathbf{x})
\frac{
\displaystyle
\int d\mathbf{x}'
\left[
H_{\alpha}(\mathbf{x},\mathbf{x}')
-
E_\theta I_{\alpha}(\mathbf{x},\mathbf{x}')
\right]
\psi_{\theta_1}(\mathbf{x}')
}{
\displaystyle
\int d\mathbf{x}\,d\mathbf{x}'\,
\psi_{\theta_1}^*(\mathbf{x})
\psi_{\theta_1}(\mathbf{x}')
I_{\alpha}(\mathbf{x},\mathbf{x}')
}
\right\}.
\label{eq:delta_energy}
\end{equation}
This expression contains a nonlocal integral over the auxiliary coordinate \(\mathbf{x}'\). In analogy with conventional VMC, we introduce a generalized local energy,
\begin{equation}
E_{\mathrm{L}}(\mathbf{x})
=
\frac{
\displaystyle
\int d\mathbf{x}'\,
H_{\alpha}(\mathbf{x},\mathbf{x}')
\psi_{\theta_1}(\mathbf{x}')
}{
\displaystyle
\int d\mathbf{x}'\,
I_{\alpha}(\mathbf{x},\mathbf{x}')
\psi_{\theta_1}(\mathbf{x}')
}=\frac{
\displaystyle
\mathbb{E}_{\mathbf{x}' \sim I_{\alpha}(\mathbf{x},\mathbf{x}')}
\frac{H_{\alpha}(\mathbf{x},\mathbf{x}')}{I_{\alpha}(\mathbf{x},\mathbf{x}')}
\psi_{\theta_1}(\mathbf{x}')
}{
\displaystyle
\mathbb{E}_{\mathbf{x}' \sim I_{\alpha}(\mathbf{x},\mathbf{x}')}
\psi_{\theta_1}(\mathbf{x}')
}.
\label{eq:generalized_local_energy}
\end{equation}
Compared with the usual local energy, this generalized local energy is nonlocal in the auxiliary coordinate space, as its evaluation requires integrating over \(\mathbf{x}'\) for each fixed \(\mathbf{x}\). In practice, this inner integral is estimated by Monte Carlo sampling, with \(\mathbf{x}'\) drawn from the Gaussian distribution proportional to \(I_{\alpha}(\mathbf{x},\mathbf{x}')\).

We further introduce the marginal sampling distribution
\begin{equation}
p_\theta(\mathbf{x})
\propto
|\psi_{\theta_1}(\mathbf{x})|
\int d\mathbf{x}'\,
I_\alpha(\mathbf{x},\mathbf{x}')
|\psi_{\theta_1}(\mathbf{x}')| .
\label{eq:marginal_sampling_distribution}
\end{equation}
The phase information omitted from this positive sampling distribution is collected into the reweighting factor
\begin{equation}
S_{\mathrm{L}}(\mathbf{x})
=
\frac{
\displaystyle
\int d\mathbf{x}'\,
I_{\alpha}(\mathbf{x},\mathbf{x}')
|\psi_{\theta_1}(\mathbf{x}')|
\operatorname{sgn}
\left[
\psi_{\theta_1}^*(\mathbf{x})
\psi_{\theta_1}(\mathbf{x}')
\right]
}{
\displaystyle
\int d\mathbf{x}'\,
I_{\alpha}(\mathbf{x},\mathbf{x}')
|\psi_{\theta_1}(\mathbf{x}')|
}
=\frac{
\displaystyle
\mathbb{E}_{\mathbf{x}' \sim I_{\alpha}(\mathbf{x},\mathbf{x}')}
|\psi_{\theta_1}(\mathbf{x}')|
\operatorname{sgn}
\left[
\psi_{\theta_1}^*(\mathbf{x})
\psi_{\theta_1}(\mathbf{x}')
\right]
}{
\displaystyle
\mathbb{E}_{\mathbf{x}' \sim I_{\alpha}(\mathbf{x},\mathbf{x}')}
|\psi_{\theta_1}(\mathbf{x}')|
}
.
\label{eq:local_sign_factor}
\end{equation}
Here, for \(z \neq 0\), \(\operatorname{sgn}(z)\) denotes the complex sign factor,
\[
\operatorname{sgn}(z) \equiv \frac{z}{|z|}.
\]
Both the numerator and denominator integrals are estimated via Monte Carlo sampling.

With these definitions, the variational energy can be written as
\begin{equation}
E_\theta
=
\frac{
\mathbb{E}_{p_\theta}
\left[
S_{\mathrm{L}}(\mathbf{x})E_{\mathrm{L}}(\mathbf{x})
\right]
}{
\mathbb{E}_{p_\theta}
\left[
S_{\mathrm{L}}(\mathbf{x})
\right]
}
.
\label{eq:energy_reweighted}
\end{equation}
Similarly, the energy variation takes the compact form
\begin{equation}
\delta E_\theta
=
2\,\mathrm{Re}
\left\{
\frac{
\mathbb{E}_{p_\theta}
\left[
\frac{\delta\psi^*(\mathbf{x})}{\psi_{\theta_1}^*(\mathbf{x})}
\left(
E_{\mathrm{L}}(\mathbf{x})-E_\theta
\right)
S_{\mathrm{L}}(\mathbf{x})
\right]
}{
\mathbb{E}_{p_\theta}
\left[
S_{\mathrm{L}}(\mathbf{x})
\right]
}
\right\}.
\label{eq:sup_energy_grad}
\end{equation}
Finally, substituting the explicit expression for the induced variation \(\delta\psi(\mathbf{x})\) into Eq.~\eqref{eq:sup_energy_grad} gives the gradient formula shown in the main text.

\subsection{Comparison between Evaluation of Expectation Value and Energy Gradient}

Comparing Eq.~\eqref{eq:mc_expectation_value} with Eq.~\eqref{eq:sup_energy_grad}, we see that the expectation value and the energy gradient are evaluated using different Monte Carlo strategies. For a generic observable, we sample the joint distribution \(q_\theta(\mathbf{x},\mathbf{x}')\) and estimate the double integral over \((\mathbf{x},\mathbf{x}')\) in a single Monte Carlo average. In contrast, for the energy gradient, we first perform the integral over \(\mathbf{x}'\), sampled from a distribution proportional to \(I_\alpha(\mathbf{x},\mathbf{x}')\), to construct the generalized local energy \(E_{\mathrm{L}}(\mathbf{x})\). The remaining integral over \(\mathbf{x}\) is then evaluated using the marginal distribution \(p_\theta(\mathbf{x})\). Below we explain the reason for using these two different strategies.

For the evaluation of observable expectation values, the one-step estimator is more efficient. If one instead used a two-step procedure, both the inner integral over \(\mathbf{x}'\) and the outer integral over \(\mathbf{x}\) would have to be estimated numerically, leading to two sources of statistical error. Sampling directly from the joint distribution \(q_\theta(\mathbf{x},\mathbf{x}')\) avoids this nested Monte Carlo structure and estimates the full double integral in a single average, thereby reducing the statistical noise.

For the energy gradient, however, the two-step procedure is essential for obtaining a stable estimator. By first integrating over \(\mathbf{x}'\), we define the generalized local energy \(E_{\mathrm{L}}(\mathbf{x})\), which plays the same stabilizing role as the local energy in conventional VMC. The exact ground state satisfies the generalized eigenvalue equation in the nonorthogonal basis,
\[
\int d\mathbf{x}'\,
H_\alpha(\mathbf{x},\mathbf{x}')
\psi_{\mathrm{GS}}(\mathbf{x}')
=
E_{\mathrm{GS}}
\int d\mathbf{x}'\,
I_\alpha(\mathbf{x},\mathbf{x}')
\psi_{\mathrm{GS}}(\mathbf{x}') .
\]
Consequently, for the exact ground state, the generalized local energy is independent of \(\mathbf{x}\) and equals \(E_{\mathrm{GS}}\). During optimization, as the variational wave function approaches the ground state, the spatial fluctuations of \(E_{\mathrm{L}}(\mathbf{x})\) are progressively suppressed. Since Eq.~\eqref{eq:sup_energy_grad} shows that the gradient is governed by the fluctuation \(E_{\mathrm{L}}(\mathbf{x})-E_\theta\), this construction naturally reduces the magnitude and variance of the gradient near convergence, thereby improving the stability of the parameter update.

The same construction also provides a stable estimator for optimizing the kernel parameter \(\alpha\). A useful consistency check is given by a plane-wave auxiliary state,
\[
\psi(\mathbf{x})=e^{i\mathbf{k}\cdot\mathbf{x}} .
\]
For this state, the basis transformation changes only the overall normalization: for any \(\alpha\), the corresponding real-space wave function takes the form
\[
\tilde{\psi}_\theta(\mathbf{r}) = C_\alpha e^{i\mathbf{k}\cdot\mathbf{r}},
\]
where \(C_\alpha\) is independent of \(\mathbf{r}\). Therefore, at fixed auxiliary wave function \(\psi(\mathbf{x})\), the variational energy is independent of \(\alpha\), and the corresponding energy derivative must vanish.

This cancellation is built into the present estimator. From Eq.~\eqref{eq:alpha_update}, the logarithmic variation associated with an infinitesimal change of \(\alpha\) is
\[
\frac{\delta\psi(\mathbf{x})}{\psi(\mathbf{x})}
=
-\frac{1}{4\alpha^2}
\frac{
\nabla_{\mathbf{x}}^2 e^{i\mathbf{k}\cdot\mathbf{x}}
}{
e^{i\mathbf{k}\cdot\mathbf{x}}
}
=
\frac{|\mathbf{k}|^2}{4\alpha^2},
\]
which is independent of \(\mathbf{x}\). Substituting this constant logarithmic variation into Eq.~\eqref{eq:sup_energy_grad} gives
\begin{equation}
\nabla_{\alpha}E_\theta
\propto
\mathrm{Re}
\left[
\frac{
\mathbb{E}_{p_\theta}
\left[
\left(E_{\mathrm{L}}(\mathbf{x})-E_\theta\right)
S_{\mathrm{L}}(\mathbf{x})
\right]
}{
\mathbb{E}_{p_\theta}
\left[
S_{\mathrm{L}}(\mathbf{x})
\right]
}
\right]
=0 .
\label{alpha_vanish}
\end{equation}
The last equality follows directly from Eq.~\eqref{eq:energy_reweighted}.
Thus, for a plane-wave auxiliary state, the \(\alpha\) derivative cancels at the level of the finite-sample reweighted estimator. This exact cancellation removes a spurious Monte Carlo contribution to the \(\alpha\) update and thereby improves the stability of kernel optimization.

\subsection{Stochastic Reconfiguration (SR) Acceleration}

After deriving the energy gradient, we consider using the popular approach for accelerating the optimization of neural quantum states known as stochastic reconfiguration (SR) to accelerate the updates~\cite{stokes2020}. In this section, we briefly introduce the idea of SR. Moreover, because SR changes accordingly after the introduction of the basis transformation, we present an analytical derivation.

SR is based on imaginary-time evolution: for any wave function \(|\tilde{\psi}\rangle\) with \(\langle\tilde{\psi}_0|\tilde{\psi}\rangle \neq 0\), the ground state \(|\tilde{\psi}_0\rangle\) satisfies
\[
|\tilde{\psi}_0\rangle = \lim_{\tau \to \infty} e^{-\tau \hat{H}} |\tilde{\psi}\rangle.
\]
We optimize a variational ansatz \(|\tilde{\psi}_\theta\rangle\) by finding parameter updates that approximate a small imaginary-time step. Setting the time step to a small \(\delta\tau\), the evolved wave function is
\begin{equation}
|\tilde{\psi}'_\theta\rangle = e^{-\delta\tau \hat{H}}|\tilde{\psi}_\theta\rangle \approx (\hat{\mathbf{1}} - \delta\tau \hat{H})|\tilde{\psi}_\theta\rangle. 
\label{eq:imag_time}
\end{equation}

Our goal is to find a parameter update \(d\theta\) such that
\(|\tilde{\psi}_{\theta+d\theta}\rangle\) approximates the imaginary-time evolved state
\(|\tilde{\psi}'_{\theta}\rangle\). This makes the variational parameter update a projection of imaginary-time evolution onto the variational manifold. We quantify this projection using the Fubini-Study distance
\begin{equation}
    D(|\tilde{\psi}\rangle,|\tilde{\phi}\rangle)
    =
    \arccos
    \frac{
    |\langle\tilde{\psi}|\tilde{\phi}\rangle|
    }{
    \|\tilde{\psi}\|\,\|\tilde{\phi}\|
    } .
    \label{eq:fubini}
\end{equation}
With the Gaussian basis transformation, the real-space wave function is written as $
    |\tilde{\psi}_{\theta}\rangle
    =
    \hat{G}_{\alpha}|\psi_{\theta_1}\rangle ,
$
where \(\hat{G}_{\alpha}\) is the Gaussian kernel operator and
\(|\psi_{\theta_1}\rangle\) denotes the auxiliary-space wave function. Expanding both
\(|\tilde{\psi}'_{\theta}\rangle\) and
\(|\tilde{\psi}_{\theta+d\theta}\rangle\) to first order, the squared Fubini-Study distance takes the second-order form
\begin{equation}
    D^2
    \left(
    |\tilde{\psi}'_{\theta}\rangle,
    |\tilde{\psi}_{\theta+d\theta}\rangle
    \right)
    =
    d\theta^{T} S d\theta
    -
    2\delta\tau\, g^{T} d\theta
    +
    C\delta\tau^2 ,
    \label{eq:fs_quadratic}
\end{equation}
where the last term is independent of \(d\theta\) and therefore does not affect the minimization. The vector \(g\) is the energy gradient, satisfying
$
    g^{T}d\theta
    =
    \delta E_{\theta}
    =
    \nabla_{\theta}E_{\theta} d\theta ,
$
with its explicit Monte Carlo estimator given in Eq.~\eqref{eq:sup_energy_grad}. The metric \(S\), i.e., the real part of the quantum geometric tensor, is in general modified by the nonorthogonal Gaussian basis. To simplify the update, we approximate the overlap operator
\[
    \hat{I}_{\alpha}
    =
    \hat{G}_{\alpha}^{\dagger}\hat{G}_{\alpha}
    \simeq
    \hat{\mathbf{1}},
\]
where \(\hat{\mathbf{1}}\) is the identity operator. This approximation corresponds to treating the Gaussian basis as approximately orthonormal and becomes exact in the real-space limit \(\alpha\to\infty\). Since the two-step optimization scheme keeps \(\alpha\) in a relatively large-\(\alpha\) regime, this approximation provides a controlled and efficient estimate of the metric. Under this approximation,
\begin{equation}
    S
    =
    \overline{O}_{\theta}^{\dagger}
    \overline{O}_{\theta},
    \qquad
    \overline{O}_{\theta}(\mathbf{x})
    =
    O_{\theta}(\mathbf{x})
    -
    \mathbb{E}_{\mathbf{x}\sim p_{\theta}}
    \left[
    O_{\theta}(\mathbf{x})
    \right],
    \label{eq:metric_approx}
\end{equation}
where \(O_{\theta}\) denotes the logarithmic derivative defined by
$
    O_{\theta}(\mathbf{x})\,d\theta
    =
    \frac{\delta\psi_{\theta_1}(\mathbf{x})}
    {\psi_{\theta_1}(\mathbf{x})}.
$
Explicitly, including both the auxiliary-wave-function parameters \(\theta_1\) and the kernel parameter \(\alpha\), we have
\begin{equation}
    O_{\theta}(\mathbf{x})
    =
    \left(
    \nabla_{\theta_1}\log\psi_{\theta_1}(\mathbf{x}),
    \;
    -\frac{1}{4\alpha^2}
    \frac{
    \nabla_{\mathbf{x}}^2\psi_{\theta_1}(\mathbf{x})
    }{
    \psi_{\theta_1}(\mathbf{x})
    }
    \right).
    \label{eq:log_derivative}
\end{equation}

With \( S \) and \( \nabla_\theta E_\theta \) known, the parameter update is obtained by solving
\[
d\theta = \arg \min_{d\theta'} \left( d\theta'^T \overline{O}_{\theta'}^\dagger \overline{O}_{\theta'} d\theta' - 2 \nabla_{\theta'} E_{\theta'} \delta \tau d\theta' \right).
\]
In practice, the number of Monte Carlo samples is typically larger than the number of parameters, making the matrix \(S\) rank-deficient and preventing direct inversion. To remedy this, Tikhonov regularization is applied. The update becomes
\[
d\theta = \left( S + \lambda I \right)^{-1} \nabla_{\theta} E_\theta,
\]
where \(\lambda\) controls the damping strength. This form is analogous to the standard SR update. Including momentum in the gradient leads to the SPRING optimization scheme~\cite{goldshlager2024}:
\begin{equation}
d\theta_t = \left( S + \lambda I \right)^{-1} \left( \nabla_{\theta_t} E_{\theta_t} + \lambda \mu \, d\theta_{t-1} \right),
\label{eq:spring}
\end{equation}
where \(\mu\) controls the momentum decay and $\theta_t$ stands for the parameters at step $t$; the case \(\mu = 0\) recovers ordinary SR.

In the first step of our two-step optimization, we optimize \(\theta_1\), which contains a large number of parameters; the quantum geometric matrix \(S\) is therefore very large. To compute \((S + \lambda I)^{-1}\) efficiently, we apply the Sherman-Morrison-Woodbury formula as described in Ref.~\cite{ren2019}. In the second step, only the single parameter \(\alpha\) is optimized, so \((S + \lambda I)^{-1}\) can be computed directly.

\subsection{Matrix Elements of Operators}

The evaluation of the energy gradient and expectation values of observables requires the matrix elements of the corresponding operators. Owing to the favorable properties of Gaussian functions, these matrix elements can be obtained analytically for most operators, which is a key motivation for choosing a Gaussian basis. Below we provide the explicit expressions for the matrix elements used in this work. For notational simplicity, we omit the explicit $\theta_1$ dependence in $\psi_{\theta_1}(\mathbf{x})$ for convenience. 

\begin{itemize}
   \item \textbf{Kinetic energy operator \(\hat{K}\).}
    The kinetic energy operator is
    \begin{equation}
        \hat{K} = -\frac{1}{2}\sum_{i=1}^{n}\nabla_{\mathbf{r}_i}^{2},
    \end{equation}
    where \(n\) is the number of electrons. A direct evaluation of the Gaussian-basis matrix element gives
    \begin{equation}
        \tilde{K}_{\alpha}(\mathbf{x},\mathbf{x}')
        =
        I_{\alpha}(\mathbf{x},\mathbf{x}')
        \frac{\alpha}{2}
        \left(
        3n-\alpha\sum_{i=1}^{n}
        |\mathbf{x}_i-\mathbf{x}'_i|^2
        \right).
        \label{eq:direct_kinetic_kernel}
    \end{equation}
    Although Eq.~\eqref{eq:direct_kinetic_kernel} is formally correct, it leads to an ill-conditioned Monte Carlo estimator for the inner integral entering the generalized local energy,
    \begin{equation}
        \int d\mathbf{x}' \,
        \tilde{K}_{\alpha}(\mathbf{x},\mathbf{x}')
        \psi(\mathbf{x}')
        =
        \mathbb{E}_{\mathbf{x'}\sim I_{\alpha}(\mathbf{x},\mathbf{x}')}
        \left[
        \frac{\alpha}{2}
        \left(
        3n-\alpha |\mathbf{x}-\mathbf{x}'|^2
        \right)
        \psi(\mathbf{x}')
        \right],
        \label{eq:kinetic_evaluation}
    \end{equation}
    where \(\mathbf{x}'\) is sampled from the normalized overlap distribution proportional to
    \(I_{\alpha}(\mathbf{x},\mathbf{x}')\), and
    \(|\mathbf{x}-\mathbf{x}'|^2\equiv \sum_i|\mathbf{x}_i-\mathbf{x}'_i|^2\). The origin of the instability can be seen from the variance of the integrand.
    For a smooth wave function, expanding around \(\mathbf{x}\) gives
    \(\psi(\mathbf{x}')=\psi(\mathbf{x})+O(|\mathbf{x}'-\mathbf{x}|)\). Keeping the leading term, one finds
    \begin{equation}
    \mathbb{E}_{\mathbf{x'}\sim I_{\alpha}(\mathbf{x}, \mathbf{x'})}
    \left[
    \frac{\alpha}{2}
    \left(
    3n-\alpha |\mathbf{x}-\mathbf{x}'|^2
    \right)
    \right]
    =0 ,
    \end{equation}
    whereas the leading contribution to the variance scales as \(O(\alpha^2)\):
    \begin{equation}
    \mathrm{Var}_{\mathbf{x'} \sim I_\alpha(\mathbf{x}, \mathbf{x'})}[g]
    \simeq
    |\psi(\mathbf{x})|^2
    \,
    \mathbb{E}_{\mathbf{x'}\sim I_{\alpha}(\mathbf{x},\mathbf{x'})}
    \left[
    \frac{\alpha^2}{4}
    \left(
    3n-\alpha |\mathbf{x}-\mathbf{x}'|^2
    \right)^2
    \right].
    \end{equation}
    Here, $
    g(\mathbf{x}')
    \equiv
    \frac{\alpha}{2}
    \left(
    3n-\alpha |\mathbf{x}-\mathbf{x}'|^2
    \right)
    \psi(\mathbf{x}')$.
    Since typical samples satisfy
    \(|\mathbf{x}-\mathbf{x}'|\sim \alpha^{-1/2}\), the factor
    \(\alpha|\mathbf{x}-\mathbf{x}'|^2\) is \(O(1)\), and hence the standard deviation of \(g\) scales as \(O(\alpha)\). Thus, at fixed sample size, the Monte Carlo standard error grows linearly with \(\alpha\). The direct estimator therefore becomes increasingly noisy in the large-\(\alpha\) limit, even though this limit should recover conventional real-space VMC.
    
    This problem can be avoided by moving the kinetic operator from the Gaussian basis function to the auxiliary wave function. Using
    \(\nabla_{\mathbf{r}}^2G_{\alpha}(\mathbf{x}',\mathbf{r})
    =
    \nabla_{\mathbf{x}'}^2G_{\alpha}(\mathbf{x}',\mathbf{r})\)
    and integrating by parts with respect to \(\mathbf{x}'\), assuming the boundary terms vanish, we obtain
    \begin{align}
        \int d\mathbf{x}' \,
        \tilde{K}_{\alpha}(\mathbf{x},\mathbf{x}')
        \psi(\mathbf{x}')
        &=
        \int d\mathbf{x}' \psi(\mathbf{x}')
        \int d\mathbf{r} \,
        G_{\alpha}(\mathbf{x},\mathbf{r})
        \left(
        -\frac{1}{2}\nabla_{\mathbf{r}}^{2}
        \right)
        G_{\alpha}(\mathbf{x}',\mathbf{r})
        \nonumber\\
        &=
        \int d\mathbf{x}' \psi(\mathbf{x}')
        \int d\mathbf{r} \,
        G_{\alpha}(\mathbf{x},\mathbf{r})
        \left(
        -\frac{1}{2}\nabla_{\mathbf{x}'}^{2}
        \right)
        G_{\alpha}(\mathbf{x}',\mathbf{r})
        \nonumber\\
        &=
        \int d\mathbf{x}' \,
        I_{\alpha}(\mathbf{x},\mathbf{x}')
        \left(
        -\frac{1}{2}\nabla_{\mathbf{x}'}^{2}
        \right)
        \psi(\mathbf{x}')
        \nonumber\\
        &=
        \mathbb{E}_{\mathbf{x'} \sim I_{\alpha}(\mathbf{x},\mathbf{x}')}
        \left[
        \left(
        -\frac{1}{2}\nabla_{\mathbf{x}'}^{2}
        \right)
        \psi(\mathbf{x}')
        \right].
        \label{eq:kinetic_ibp}
    \end{align}
    Equivalently, the kinetic contribution may be represented by the operator-valued kernel
    \begin{equation}
        K_{\alpha}(\mathbf{x},\mathbf{x}')
        =
        I_{\alpha}(\mathbf{x},\mathbf{x}')
        \left(
        -\frac{1}{2}\nabla_{\mathbf{x}'}^{2}
        \right).
        \label{eq:kinetic}
    \end{equation}
    
    The advantage of Eq.~\eqref{eq:kinetic} is that it removes the large, fluctuating prefactor present in Eq.~\eqref{eq:kinetic_evaluation}. In the limit \(\alpha\to\infty\), the overlap distribution becomes sharply peaked around \(\mathbf{x}'=\mathbf{x}\), and Eq.~\eqref{eq:kinetic_ibp} reduces smoothly to the conventional VMC kinetic estimator
    $
        \left(
        -\frac{1}{2}\nabla_{\mathbf{x}}^{2}
        \right)\psi(\mathbf{x}).
    $
    This integration-by-parts form therefore resolves the variance divergence of the direct kinetic-energy estimator.

    \item \textbf{Coulomb interaction operator $\hat{C}$}.

    For the three-dimensional homogeneous electron gas, the Coulomb interaction operator is given by
    \begin{equation}
        \hat{C} = \frac{1}{2}\sum_{i,j}{}' \sum_{\mathbf{L}_s} \frac{1}{|\mathbf{r}_i - \mathbf{r}_j + \mathbf{L}_s|},
    \end{equation}
    where $\mathbf{L}_s$ are lattice vectors of the simulation cell under periodic boundary conditions (PBC). The sum over $\mathbf{L}_s$ enforces PBC, and the prime on $\sum_{i,j}'$ indicates that terms with $i=j$ are omitted when $\mathbf{L}_s = 0$. Using the Ewald summation technique \cite{ewald1921}, the long-range Coulomb interaction is split into short-range and long-range parts via \(
        \frac{1}{r} = \frac{\operatorname{erfc}(\omega r)}{r} + \frac{\operatorname{erf}(\omega r)}{r}
    \), where $\operatorname{erfc}(x)=1-\operatorname{erf}(x)$ and $\operatorname{erf}(x)=\frac{2}{\sqrt{\pi}}\int_0^x e^{-t^2}dt$. The parameter $\omega$ controls the convergence rates in real and reciprocal space. The short-range part is summed in real space, while the long-range part is handled in reciprocal space, yielding
    \begin{equation}
        \begin{aligned}
            \hat{C} &= \frac{1}{2}\sum_{i,j}{}' \sum_{\mathbf{L}} \frac{\operatorname{erfc}(\omega|\mathbf{r}_i - \mathbf{r}_j + \mathbf{L}|)}{|\mathbf{r}_i - \mathbf{r}_j + \mathbf{L}|} \\
            &\quad + \frac{1}{2\Omega}\sum_{i,j}\sum_{\mathbf{G}\neq 0} \frac{4\pi}{|\mathbf{G}|^2} \exp\!\left(-\frac{|\mathbf{G}|^2}{4\omega^2}\right) \exp\!\bigl(i\mathbf{G}\cdot(\mathbf{r}_i - \mathbf{r}_j)\bigr) \\
            &\quad - \frac{N\omega}{\sqrt{\pi}} - \frac{\pi N^2}{2\Omega\omega^2},
        \end{aligned}
        \label{eq:ewald}
    \end{equation}
    where $\Omega$ is the cell volume. The matrix elements of this operator defined in Eq.~\eqref{eq:ewald} can also be derived analytically \cite{ahlrichs2006}. The final result has the form:
    \begin{equation}
        \begin{aligned}
            \frac{C_{\alpha}}{I_{\alpha}}(\mathbf{x},\mathbf{x}') &=
            \sum_{j\neq k}\sum_{\mathbf{L}_s}\sqrt{\frac{\alpha}{\pi}}\Biggl[
             F_0\!\left(\alpha|\boldsymbol{\xi}_j-\boldsymbol{\xi}_k+\mathbf{L}_s|^2\right) - \frac{\omega}{\sqrt{\omega^2+\alpha}} F_0\!\left(\frac{\omega^2\alpha|\boldsymbol{\xi}_j-\boldsymbol{\xi}_k+\mathbf{L}_s|^2}{\omega^2+\alpha}\right) \Biggr] \\
            &\quad + \frac{1}{2}\sum_{j}\sum_{\mathbf{L}_s\neq 0} \left( \frac{1}{|\mathbf{L}_s|} - \frac{2\omega}{\sqrt{\pi}} F_0(\omega^2|\mathbf{L}_s|^2) \right) \\
            &\quad + \frac{1}{2\Omega}\sum_{\mathbf{G}\neq 0} \frac{4\pi}{|\mathbf{G}|^2} \exp\!\left(-\frac{|\mathbf{G}|^2}{4\omega^2} - \frac{|\mathbf{G}|^2}{4\alpha}\right) \left| \sum_j e^{i \boldsymbol{\xi}_j \cdot \mathbf{G}} \right|^2 \\
            &\quad + \frac{N}{2\Omega}\sum_{\mathbf{G}\neq 0} \frac{4\pi}{|\mathbf{G}|^2} \exp\!\left(-\frac{|\mathbf{G}|^2}{4\omega^2}\right) \left(1 - \exp\!\left(-\frac{|\mathbf{G}|^2}{4\alpha}\right)\right) \\
            &\quad - \frac{N\omega}{\sqrt{\pi}} - \frac{\pi N^2}{2\Omega \omega^2}.
        \end{aligned}
    \end{equation}
    Here $\boldsymbol{\xi}_i \equiv \frac{\mathbf{x}_i+\mathbf{x}'_i}{2}$, and $F_m(x)$ denotes the Boys function (incomplete gamma function), defined as
    \begin{equation}
        F_m(x) = \int_0^1 \exp(-x t^2)\, t^{2m} dt.
    \end{equation}
    This function can be evaluated efficiently using Padé approximants (ratio of two polynomials). In this work we employ the coefficients given by Schaad in 1971 \cite{schaad1971}, which yield an error smaller than $3\times10^{-9}$.

    \item \textbf{Correlation function operators}.

    To characterize correlation effects in the variational wave function, we introduce the spin-resolved two-body density operator
\begin{equation}
    \hat{\rho}_{mn}^{(2)}(\boldsymbol{\tau}_1,\boldsymbol{\tau}_2)
    =
    \frac{1}{N(N-1)}
    \sum_{i\neq j}
    \delta(\boldsymbol{\tau}_1-\mathbf{r}_i)
    \delta(\boldsymbol{\tau}_2-\mathbf{r}_j)
    \delta_{m,s_i}\delta_{n,s_j},
\end{equation}
where \(N\) is the total number of electrons, \(\mathbf{r}_i\) and \(s_i\in\{\uparrow,\downarrow\}\) denote the position and spin of the \(i\)-th electron, respectively, and \(m,n\) are spin indices. The pair correlation function, the spin-resolved pair correlation function, and the static structure factor are defined as
\begin{equation}
    \begin{aligned}
        \hat{g}(\boldsymbol{\tau})
        &=
        \Omega
        \int d\boldsymbol{\tau}_1
        \sum_{m,n}
        \hat{\rho}_{mn}^{(2)}
        (\boldsymbol{\tau}_1,\boldsymbol{\tau}_1+\boldsymbol{\tau}),
        \\[4pt]
        \hat{g}_s(\boldsymbol{\tau})
        &=
        \Omega
        \int d\boldsymbol{\tau}_1
        \sum_{m,n}
        (2\delta_{mn}-1)
        \hat{\rho}_{mn}^{(2)}
        (\boldsymbol{\tau}_1,\boldsymbol{\tau}_1+\boldsymbol{\tau}),
        \\[4pt]
        \hat{S}(\mathbf{k})
        &=
        1-N\delta_{\mathbf{k},0}
        +(N-1)
        \int d\boldsymbol{\tau}_1 d\boldsymbol{\tau}
        \sum_{m,n}
        \hat{\rho}_{mn}^{(2)}
        (\boldsymbol{\tau}_1,\boldsymbol{\tau}_1+\boldsymbol{\tau})
        e^{-i\mathbf{k}\cdot\boldsymbol{\tau}} .
    \end{aligned}
\end{equation}
Here, the sums over \(m,n\) run over
\(\{\uparrow,\downarrow\}\), and \(\Omega\) is the volume of the simulation cell.
The normalized matrix element of the static structure factor in the Gaussian basis is
\begin{equation}
    \frac{S_\alpha(\mathbf{k})}{I_{\alpha}}(\mathbf{x},\mathbf{x}')
    =
    1-N\delta_{\mathbf{k},0}
    +
    \frac{1}{N}
    e^{-\mathbf{k}^2/(4\alpha)}
    \sum_{i\neq j}
    e^{i\mathbf{k}\cdot(\boldsymbol{\xi}_i-\boldsymbol{\xi}_j)},
    \qquad
    \boldsymbol{\xi}_i
    \equiv
    \frac{\mathbf{x}_i+\mathbf{x}'_i}{2}.
\end{equation}
This expression is smooth and can be evaluated accurately by Monte Carlo sampling. After obtaining
\(\langle \hat{S}(\mathbf{k})\rangle\), the pair correlation function follows from the inverse Fourier relation
\begin{equation}
    \langle \hat{g}(\boldsymbol{\tau}) \rangle
    =
    1+
    \frac{1}{N}
    \sum_{\mathbf{k}}
    e^{i\mathbf{k}\cdot\boldsymbol{\tau}}
    \left[
    \langle \hat{S}(\mathbf{k}) \rangle -1
    \right].
\end{equation}
    
\end{itemize}

\section{Computational details}
In this section, we provide the specific numerical details, including network architectural details, sampling details and training details.

\subsection{Wavefunction Architectural Details}

To assess the energy improvement brought by the basis transformation, we test two different architectures for the auxiliary-space wave function \(\psi_{\theta_1}(\mathbf{x})\): FermiNet and a message-passing neural network (MPNN). For each architecture, we further vary the number of trainable parameters, or equivalently the network complexity. In this section, we provide the detailed network specifications used in our calculations.

For FermiNet, the electronic wave function is written as~\cite{pfau2020}
\begin{equation}
\begin{aligned}
\psi_{\theta_1}
(&\mathbf{r}_1^\uparrow,\ldots,\mathbf{r}_{n_\uparrow}^\uparrow;
\mathbf{r}_1^\downarrow,\ldots,\mathbf{r}_{n_\downarrow}^\downarrow)
\\
&=
\sum_{k=1}^{n_{\mathrm{det}}}
\omega_k
\det
\left[
\phi_i^{k\uparrow}
\left(
\mathbf{r}_j^\uparrow;
\{\mathbf{r}_{/j}^\uparrow\},
\{\mathbf{r}^\downarrow\}
\right)
\right]
\det
\left[
\phi_i^{k\downarrow}
\left(
\mathbf{r}_j^\downarrow;
\{\mathbf{r}^\uparrow\},
\{\mathbf{r}_{/j}^\downarrow\}
\right)
\right],
\end{aligned}
\label{eq:ferminet}
\end{equation}
where \(\phi_i^{k\uparrow}\) and \(\phi_i^{k\downarrow}\) are many-body orbitals represented by neural networks. The wave function is expressed as a linear combination of products of spin-up and spin-down determinants. The number of determinants \(n_{\mathrm{det}}\) controls the expressive power of the ansatz, with larger values providing a more flexible variational form. For the homogeneous electron gas (HEG), periodic boundary conditions are required. We therefore adopt the periodic FermiNet construction proposed in Ref.~\cite{cassella2023}, which incorporates periodic input embeddings and a periodic multiplicative envelope. In our calculations, we test FermiNet with \(n_{\mathrm{det}}=1,4,16\).

For MPNN, we follow the message-passing neural quantum state ansatz introduced in Ref.~\cite{pescia2024}. The wave function takes the determinant form
\begin{equation}
\psi_{\theta_1}(\mathbf{x})
=
\det
\left[
\varphi_\mu(\mathbf{y}_i(\mathbf{x}))
\right],
\label{eq:mpnn_wavefunction}
\end{equation}
where
\begin{equation}
\varphi_\mu(\mathbf{y}_i)
=
\exp[J(\mathbf{Y},\mu)]
\phi_\mu(\mathbf{y}_i).
\label{eq:mpnn}
\end{equation}
Here \(J\) and \(\mathbf{Y}=\{\mathbf{y}_i\}\) are many-body neural-network functions generated by the message-passing architecture, while \(\phi_\mu\) denotes a reference orbital. The complexity of the MPNN ansatz is controlled by the number of message-passing iterations: increasing the number of iterations increases the expressiveness of the network. To stabilize the optimization, we also include skip connections and layer normalization, following Ref.~\cite{smith2024}. In this work, we use one or two message-passing iterations.

The choice of reference orbitals \(\phi_\mu\) depends on the physical regime of interest. For the HEG, we consider two types of reference states. The first is the plane-wave reference,
\begin{equation}
\phi_{\mathbf{k}}(\mathbf{x})
=
\exp(i\mathbf{k}\cdot\mathbf{x}),
\qquad
\mathbf{k}
=
\frac{2\pi}{L}\mathbf{n},
\end{equation}
which is a natural choice for describing the Fermi-liquid regime. The second is a set of localized Gaussian orbitals centered at body-centered-cubic (BCC) lattice sites \(\mathbf{R}_\mu\),
\begin{equation}
\phi_{\mu}(\mathbf{r},s)
=
\sum_{\mathbf{R}_n}
\exp
\left[
-\beta
|\mathbf{r}-\mathbf{R}_\mu+\mathbf{R}_n|^2
\right]
\delta_{s,s_\mu},
\end{equation}
which is designed to capture the Wigner-crystal regime. Here \(\mu=(\mathbf{R}_\mu,s_\mu)\), \(\beta\) is a variational parameter, and the sum over simulation-cell lattice vectors \(\mathbf{R}_n\) enforces periodicity.

\begin{table}[ht]
\caption{Number of trainable parameters \(N_{\mathrm{params}}\) for different network architectures and hyperparameters. For FermiNet (FN), we use \(n_{\mathrm{det}}=1,4,16\). For MPNN, we employ both plane-wave (PW) and Gaussian-orbital (GO) reference states with one or two message-passing iterations.}
\label{tab:supp_parameters_number}
\renewcommand{\arraystretch}{1.2}
\begin{tabular}{c | c | l}
\hline\hline
Architecture & Hyperparameter & \(N_{\mathrm{params}}\) \\[0.3ex]
\hline
\(\mathrm{FN}\)      & \(n_{\mathrm{det}}=1\)  & \(628\,850\) \\[0.3ex]
\(\mathrm{FN}\)      & \(n_{\mathrm{det}}=4\)  & \(640\,232\) \\[0.3ex]
\(\mathrm{FN}\)      & \(n_{\mathrm{det}}=16\) & \(685\,760\) \\[0.3ex]
\(\mathrm{MPNN\text{-}PW}\) & iteration \(=1\) & \(10\,689\) \\[0.3ex]
\(\mathrm{MPNN\text{-}PW}\) & iteration \(=2\) & \(22\,297\) \\[0.3ex]
\(\mathrm{MPNN\text{-}GO}\) & iteration \(=2\) & \(21\,555\) \\[0.3ex]
\hline\hline
\end{tabular}
\end{table}

The number of trainable parameters for the architectures and hyperparameters used in this work is summarized in Table~\ref{tab:supp_parameters_number}. For FermiNet, the network complexity is controlled by \(n_{\mathrm{det}}\), while for MPNN it is controlled by the number of message-passing iterations. Increasing these architectural hyperparameters leads to a substantial increase in the number of trainable parameters. By contrast, the basis transformation introduces only one additional variational parameter, namely the Gaussian width \(\alpha\).

\subsection{Sampling and Training Details}

The evaluation of energy gradients and expectation values requires sampling over \(\mathbf{x}\) and \(\mathbf{x}'\). Because the conditional distribution of \(\mathbf{x}'\) is Gaussian, it can be sampled directly. Sampling of \(\mathbf{x}\) is performed using MCMC. In each MCMC step, a new configuration \(\tilde{\mathbf{x}}\) is proposed by adding Gaussian noise to the current one:
\[
\tilde{\mathbf{x}} = \mathbf{x} + \tau\,\mathcal{N}(0,1),
\]
where the step size \(\tau\) is dynamically adjusted by a factor of 1.1 or 0.9 every 100 optimization steps to keep the average acceptance rate near 0.50. Between each optimization step, we run 20 MCMC steps to reduce autocorrelation, and only the final electron configurations are used for optimization.

In practice, the number of parallel Markov chains is chosen depending on the system and the parameter update step. Initial electron configurations are chosen uniformly at random inside the simulation cell. Before optimization, we thermalize the samples with 10,000 MCMC steps. The parameters \(\theta_1\) are initialized randomly.

We optimize both \(\theta = (\theta_1, \alpha)\) using SPRING, as defined in Eq.~\eqref{eq:spring}. Following the stabilization strategy of Ref.~\cite{goldshlager2024}, we further constrain the update \(d\theta_t\) by rescaling it whenever its Euclidean norm exceeds a threshold \(C\). The learning rate \(\eta\) follows an inverse-decay schedule,
\[
\eta
=
\eta_0
\left(
1+\frac{t}{T}
\right)^{-1},
\]
where \(\eta_0\) is the initial learning rate and \(T\) is the number of optimization steps after which the learning rate is reduced by a factor of two. The parameters are updated according to
\[
\theta_{t+1} =  \theta_{t} - \eta \, d\theta_{t}
\]

For parameter training, we employ a two-step optimization strategy. The details of each optimization step are as follows.

\textbf{Training details for step I optimization.} In step I, we fix the basis parameter at the real-space limit, \(\alpha\to\infty\). In practice, we set \(\alpha=1\times10^{20}\), for which the Gaussian basis effectively reduces to the real-space basis. Step I is therefore equivalent to conventional VMC optimization, and our choice of training hyperparameters largely follows those used in previous works~\cite{cassella2023, smith2024}.

For the FermiNet calculations, we consider an unpolarized 14-electron system in a cubic simulation cell and run 1024 Markov chains in parallel. For the MPNN calculations, we consider an unpolarized 36-electron system in a rectangular simulation cell with aspect ratio \(3:3:2\), and run 512 parallel Markov chains.

For FermiNet, we use
\[
\lambda=0.001,
\qquad
\mu=0.99,
\qquad
T=20000,
\qquad
C=\frac{0.001}{\eta},
\]
with initial learning rate \(\eta_0=0.05\). For MPNN, we use
\[
\lambda=0.001,
\qquad
\mu=0.9,
\qquad
T=2000.
\]
The initial learning rate \(\eta_0\) and the norm constraint \(C\) are chosen to depend on the density parameter \(r_s\), in order to account for the different energy scales. For \(r_s=5,10,30\), we use \(C=1\) and set \(\eta_0=0.1,0.5,1.0\), respectively. For \(r_s=50,80,100\), we set \(C=\eta_0=2,3,5\), respectively.

For the FermiNet architecture, the large number of parameters can lead to undertraining if the learning rate decays too early, before the parameters reach a sufficiently low-energy region. To mitigate this issue, we use a two-stage optimization procedure. The parameters obtained from the first optimization stage are used as the initial parameters for a second optimization of \(\theta_1\), in which the learning rate is reset to its initial value and then decayed again according to the same schedule. This procedure alleviates premature learning-rate decay. For most calculations, the second stage is run for 50,000 epochs.

\textbf{Training details for step II optimization.} 
In step II, we fix the wave-function parameters \(\theta_1\) obtained from step I and optimize only the basis parameter \(\alpha\). Unlike conventional VMC, the basis transformation introduces an additional Monte Carlo integral over \(\mathbf{x}'\). Let \(N'\) denote the number of \(\mathbf{x}'\) samples and \(N\) the number of \(\mathbf{x}\) samples, or equivalently the number of parallel MCMC chains. The computational cost of evaluating the energy gradient in each epoch then scales as \(O(NN')\), which is substantially higher than the \(O(N)\) cost of conventional VMC. To keep the computational time under control, we reduce \(N\) in step II. This reduction is feasible because only a single parameter, \(\alpha\), is optimized in this step, making the parameter update much less demanding than the high-dimensional optimization of \(\theta_1\). The additional gradient noise introduced by using fewer MCMC samples is therefore acceptable.

The number of inner samples \(N'\) can also be kept moderate. The reason is that the two-step strategy prevents \(\alpha\) from becoming too small. In the limit \(\alpha\to\infty\), the overlap kernel becomes sharply localized,
\[
I_\alpha(\mathbf{x},\mathbf{x}')
\to
\delta(\mathbf{x}-\mathbf{x}'),
\]
so that \(\mathbf{x}'\) is constrained to \(\mathbf{x}'=\mathbf{x}\), and the integral over \(\mathbf{x}'\) can be evaluated exactly with \(N'=1\). For large but finite \(\alpha\), the distribution of \(\mathbf{x}'\) remains sufficiently localized around \(\mathbf{x}\), so a modest value of \(N'\) is adequate.

Because both \(N\) and \(N'\) can be kept small in step II, the per-epoch cost remains controlled. In addition, single-parameter optimization converges faster and more stably, so the total number of epochs required for convergence is also limited. In practice, for FermiNet calculations on the 14-electron system, we use 32 parallel Markov chains and independently sample 100 values of \(\mathbf{x}'\) for each \(\mathbf{x}\). For MPNN calculations on the 36-electron system, we use 16 parallel chains and independently sample 40 values of \(\mathbf{x}'\) for each \(\mathbf{x}\). With these settings, the per-epoch computational time is comparable to that of step I. Since convergence in step II typically requires fewer epochs, about 1000 epochs in our calculations, the overall runtime is significantly shorter than that of step I.

Since only one parameter is optimized in step II, the quantum geometric tensor \(S\) reduces to a scalar. In this case, SR differs from direct gradient descent only by a multiplicative factor, which effectively provides an adaptive learning-rate rescaling and stabilizes the optimization without problem-specific tuning. 

In principle, the initial value of \(\alpha\) should correspond to the limit \(\alpha_0\to\infty\). In practice, however, it is sufficient to start from a large but finite \(\alpha_0\). The choice of \(\alpha_0\) has a substantial effect on the convergence rate, since an excessively large value requires many training epochs before \(\alpha\) reaches the relevant regime. We therefore first probe several values of \(\alpha_0\), separated by orders of magnitude, using a larger initial learning rate \(\eta_0'\). For example, for FermiNet with \(n_{\mathrm{det}}=1\) at \(r_s=5\), we find that starting from \(\alpha_0=10^3\) leads to a monotonic decrease of \(\alpha\) throughout the probing run. Reducing the initial value to \(\alpha_0=10^2\), we observe that \(\alpha\) begins to oscillate after reaching approximately \(2\times10^1\). Based on this observation, we choose \(\alpha_0=3\times10^1\) for the final Step-II training and reduce the initial learning rate from \(\eta_0'\) to \(\eta_0\). In our calculations, \(\eta_0'/\eta_0\) is typically between 2 and 10.

Owing to the stability of single-parameter optimization, the results are insensitive to the detailed choices of the SPRING hyperparameters. In practice, for both FermiNet and MPNN, we set
\[
\lambda = 0.001,
\qquad
\mu = 0.99 .
\]
For most calculations, the learning rate \(\eta\) is kept constant in the range \(0.2\text{--}0.4\), and the norm constraint \(C\) is chosen between \(1\times10^{-6}\) and \(4\times10^{-6}\).
The exceptions are the high-density cases \(r_s=1\) and \(r_s=2\), for which the optimization is more challenging. As \(r_s\) decreases, the kinetic energy becomes increasingly dominant and the optimized state approaches the free-electron limit. In this regime, the \(\alpha\) gradient is strongly suppressed, as discussed around Eq.~\eqref{alpha_vanish}, and a larger update scale is needed to obtain efficient convergence. We therefore use larger learning rates, \(\eta=100\) for \(r_s=1\) and \(\eta=10\) for \(r_s=2\), together with increased norm constraints, \(C=1\times10^{-3}\) and \(C=4\times10^{-5}\), respectively. We also set \(\mu=0\) in these high-density calculations, which we find gives more stable updates.

\section{Additional numerical results}

\subsection{Effectiveness of Two-Step Optimization}

\begin{figure}[t]
\centering
\includegraphics[width=0.85\linewidth]{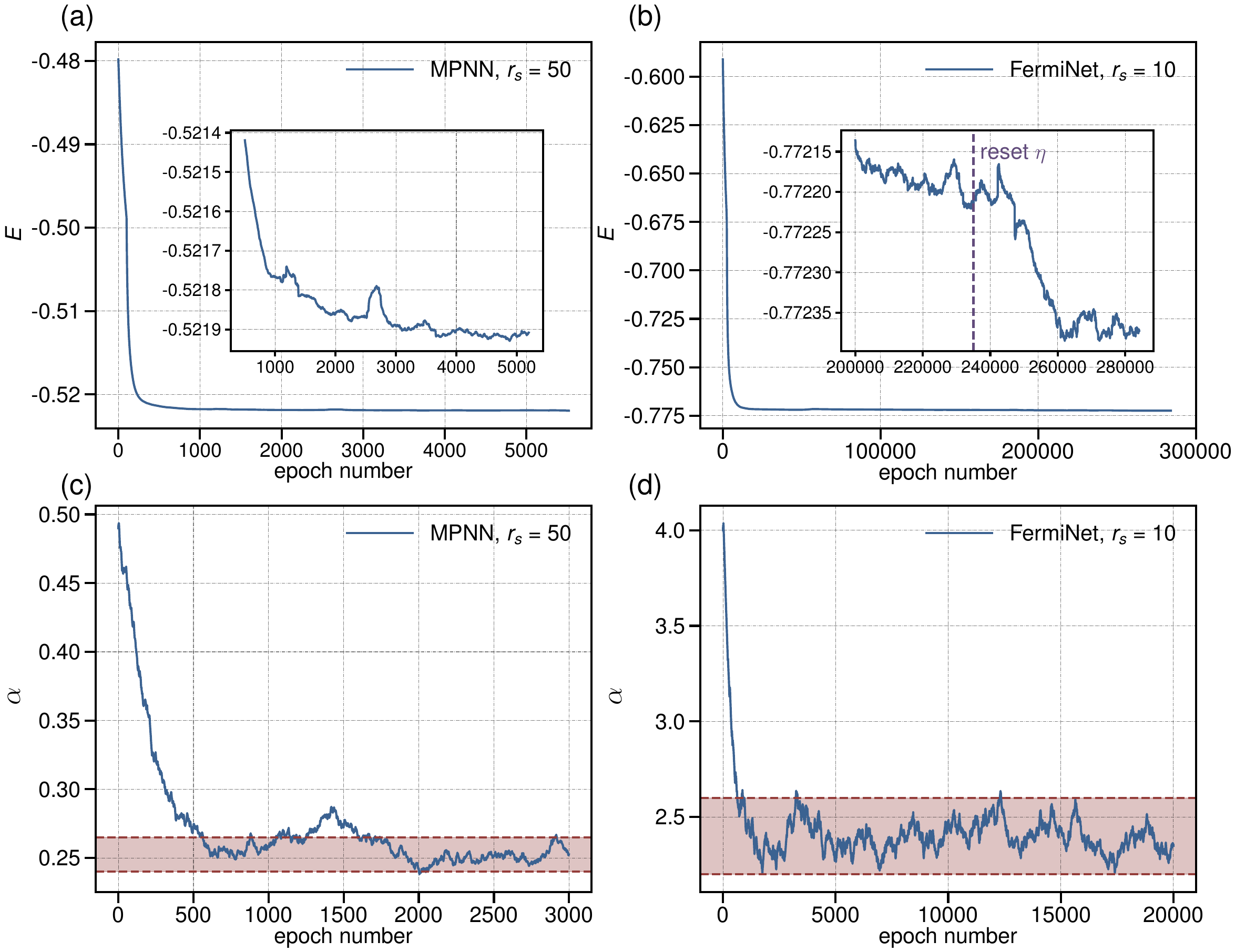}
\caption{Evolution of the energy and \(\alpha\) with epoch number during the two-step update. Panels (a) and (b) show the energy as a function of epoch in Step I for MPNN with 36 electrons (\(r_s = 50\), two iterations, PW reference state) and FermiNet with 14 electrons (\(r_s = 10\), \(n_{\text{det}} = 16\)), respectively. Panels (c) and (d) show the evolution of \(\alpha\) with epoch in Step II for the same two systems.}
\label{fig:two_step}
\end{figure}

To illustrate the effectiveness of the two-step optimization strategy, we present representative optimization trajectories for both the energy and the basis parameter. We consider two cases: an MPNN ansatz with 36 electrons at \(r_s=50\), using two message-passing iterations and the plane-wave reference state, and a FermiNet ansatz with 14 electrons at \(r_s=10\) and \(n_{\mathrm{det}}=16\). The results are summarized in Fig.~\ref{fig:two_step}. Figures~\ref{fig:two_step}(a) and~\ref{fig:two_step}(b) show the energy evolution during step I, where only the wave-function parameters \(\theta_1\) are optimized. In both cases, the energy decreases rapidly at the beginning and then gradually saturates, leaving only residual fluctuations as the optimization converges. In step II, we fix \(\theta_1\) at the values obtained from step I and optimize only the basis parameter \(\alpha\), initialized from a relatively large value \(\alpha_0\). As shown in Figs.~\ref{fig:two_step}(c) and~\ref{fig:two_step}(d), \(\alpha\) first decreases sharply and then stabilizes within a finite range, indicating that the single-parameter optimization is stable.

\begin{figure}[t]
\centering
\includegraphics[width=0.8\linewidth]{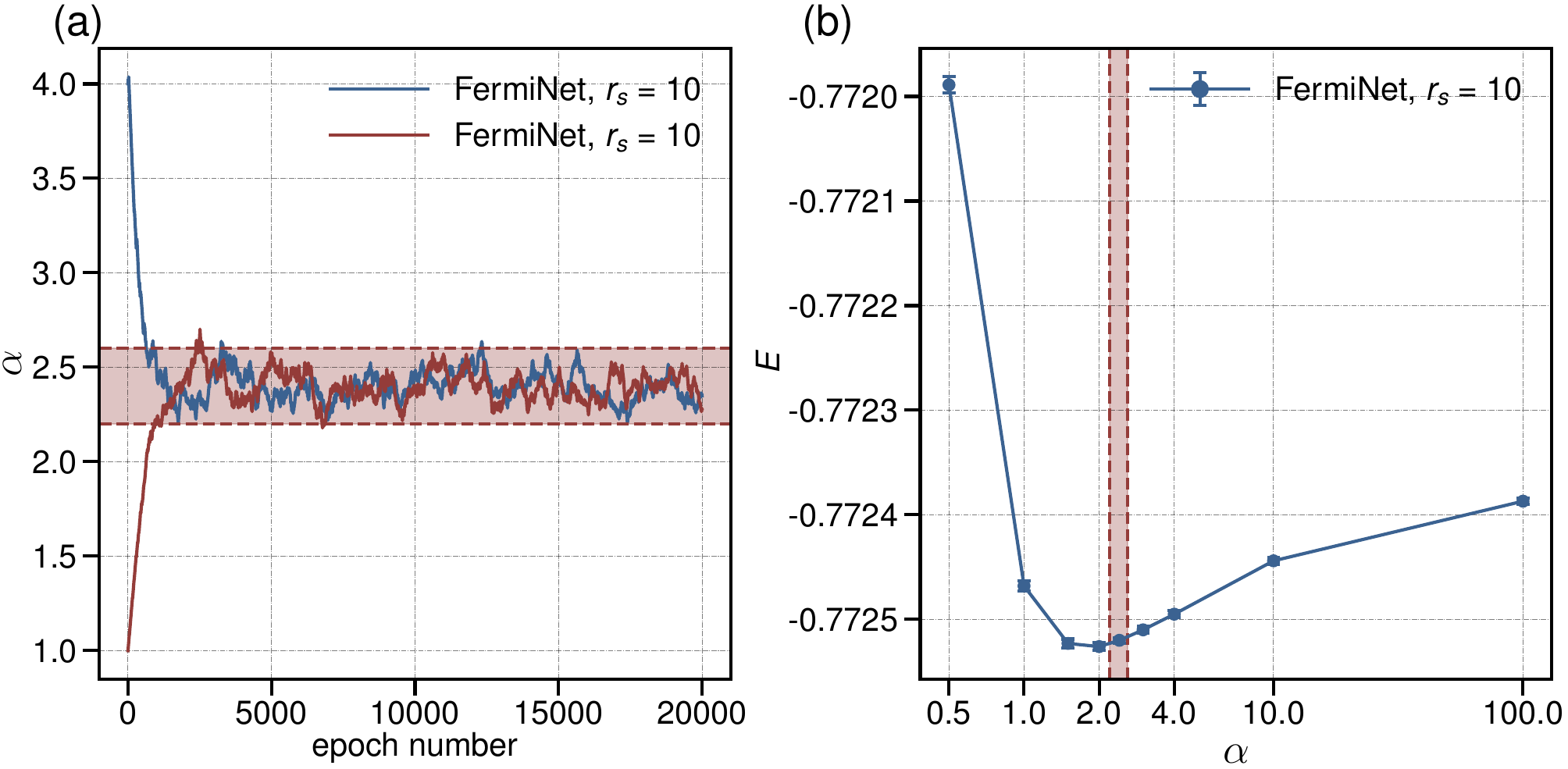}
\caption{Update of \(\alpha\) and the energy as a function of \(\alpha\) for the FermiNet system with 14 electrons, \(r_s = 10\), and \(n_{\text{det}} = 16\). Panel (a) shows the evolution of \(\alpha\) starting from two different initial values, \(\alpha_0 = 1\) (small) and \(\alpha_0 = 4\) (large). Panel (b) shows the energy \(E\) as a function of \(\alpha\).}
\label{fig:alpha_scan}
\end{figure}

To directly assess the effect of \(\alpha\) on the total energy, we examine the energy \(E\) as a function of \(\alpha\). As noted earlier, after the basis transformation the energy gradient requires separate Monte Carlo sampling over \(\mathbf{x}\) and \(\mathbf{x}'\); this introduces additional statistical error due to finite sampling of \(\mathbf{x}'\), making the energy expectation values evaluated at each epoch less reliable. Therefore, we cannot directly obtain a smooth curve of \(E\) versus epoch number. Instead, because only a single parameter \(\alpha\) is involved, we discretely sample a set of \(\alpha\) values near the stable region and compute the corresponding energies \(E\) using a single Monte Carlo integration (over both \(\mathbf{x}\) and \(\mathbf{x}'\)). Taking the example of FermiNet with 14 electrons, \(r_s = 10\), and \(n_{\text{det}} = 16\), Fig.~\ref{fig:alpha_scan} shows the update of \(\alpha\) and the energy \(E\) as a function of \(\alpha\). In Fig.~\ref{fig:alpha_scan}(a), we initialize \(\alpha\) from two different values, \(\alpha_0 = 1\) and \(\alpha_0 = 4\). In both cases, \(\alpha\) converges to a range between 2.2 and 2.6. Correspondingly, Fig.~\ref{fig:alpha_scan}(b) shows the energy \(E\) as a function of \(\alpha\); the range obtained from the \(\alpha\) update coincides with the minimum of the energy–\(\alpha\) curve, confirming the effectiveness of the \(\alpha\) update.

\subsection{Failure of Further Optimization}
In the two-step optimization protocol, we first fix \(\alpha\to\infty\) and optimize the wave-function parameters \(\theta_1\), and then fix \(\theta_1\) and optimize the basis parameter \(\alpha\). A natural extension is to add a third step, in which \(\alpha\) is fixed at the optimized value \(\alpha^*\) obtained from step II and \(\theta_1\) is optimized again. This additional step is, however, computationally expensive. Since the wave-function network contains a large number of parameters, stable optimization of \(\theta_1\) requires a large number of \(\mathbf{x}\) samples, comparable to that used in step I. At the same time, after step II the optimized \(\alpha^*\) is finite, so an accurate evaluation of the inner integral over \(\mathbf{x}'\) requires a sufficiently large number \(N'\) of \(\mathbf{x}'\) samples. Because the cost of evaluating the energy gradient scales as \(O(NN')\), the per-epoch cost of this third step becomes prohibitively high.

Moreover, we find that the energy improvement obtained from step III is not robust. To illustrate this, we tested step III using the FermiNet architecture for the unpolarized 14-electron system at \(r_s=10\), with different values of \(n_{\mathrm{det}}\). We used 1024 parallel Markov chains and independently sampled 100 values of \(\mathbf{x}'\) for each \(\mathbf{x}\), while keeping the remaining hyperparameters the same as in step I. The results are summarized in Table~\ref{tab:supp_step_3}. Step III does not consistently lower the energy: for \(n_{\mathrm{det}}=1\), no improvement is observed; for \(n_{\mathrm{det}}=4\), the energy decreases slightly; and for \(n_{\mathrm{det}}=16\), the energy increases. 

This failure of step III has two possible explanations. First, the large computational cost prevents us from using a sufficiently large \(N'\), which can lead to noisy or biased estimates of the energy gradient and thereby degrade the update of \(\theta_1\). Second, the parameters optimized by the two-step procedure are already well learned, leaving no room for further improvement; consequently, the energy starts to oscillate as the number of training epochs increases.

\begin{table}[ht]
\caption{Total energies, in Hartree, for the unpolarized $N=14$ electron system using the FermiNet architecture at $r_s = 10$. Here $E'$ is the energy before basis transformation (Step I only), $E$ is the energy after basis transformation (including Step II), and $E''$ is the energy obtained after the additional Step III optimization.}
\label{tab:supp_step_3}
\begin{center}
\renewcommand{\arraystretch}{1.4}
\begin{tabular}{r | l l l}
\hline\hline
$n_{\text{det}}$ & $E'$ & $E$& $E''$  \\[0.3ex]
\hline
1   & -0.770977(3)  & -0.771252(3)  & -0.771250(4)   \\[0.3ex]
4   & -0.771216(4)  & -0.771452(4)  & -0.771521(4)   \\[0.3ex]
16  & -0.772378(3)  & -0.772519(3)  & -0.772501(4)    \\
\hline\hline
\end{tabular}
\end{center}
\end{table}

\subsection{Electron-Electron Correlations after Basis Transformation}

\begin{figure}[t]
\centering
\includegraphics[width=1.0\linewidth]{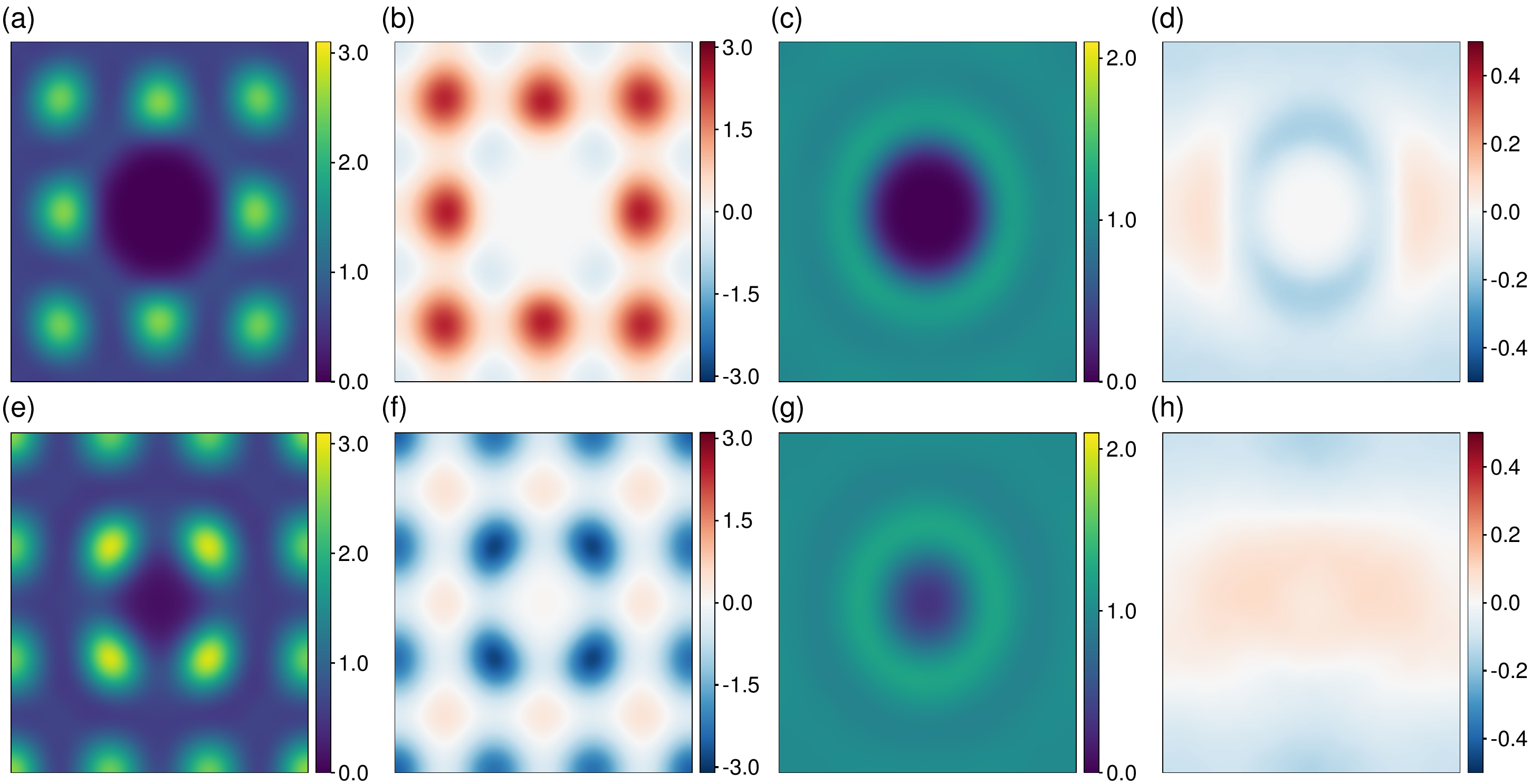}
\caption{Spatial distributions of the pair correlation function \(g(\mathbf{r})\) and the spin correlation function \(g_s(\mathbf{r})\). Panels (a) and (b) show \(g(\mathbf{r})\) and \(g_s(\mathbf{r})\) in the \(z=0\) plane for \(r_s=87\). Panels (c) and (d) show the same quantities in the \(z=0\) plane for \(r_s=50\). Panels (e) and (f) show \(g(\mathbf{r})\) and \(g_s(\mathbf{r})\) in the \(z=a_{\mathrm{BCC}}/2\) plane for \(r_s=87\), where \(a_{\mathrm{BCC}}\) is the lattice constant of the BCC unit cell. Panels (g) and (h) show the same in the \(z=a_{\mathrm{BCC}}/2\) plane for \(r_s=50\).}
\label{fig:real_correlation}
\end{figure}

\begin{figure}[t]
\centering
\includegraphics[width=0.6\linewidth]{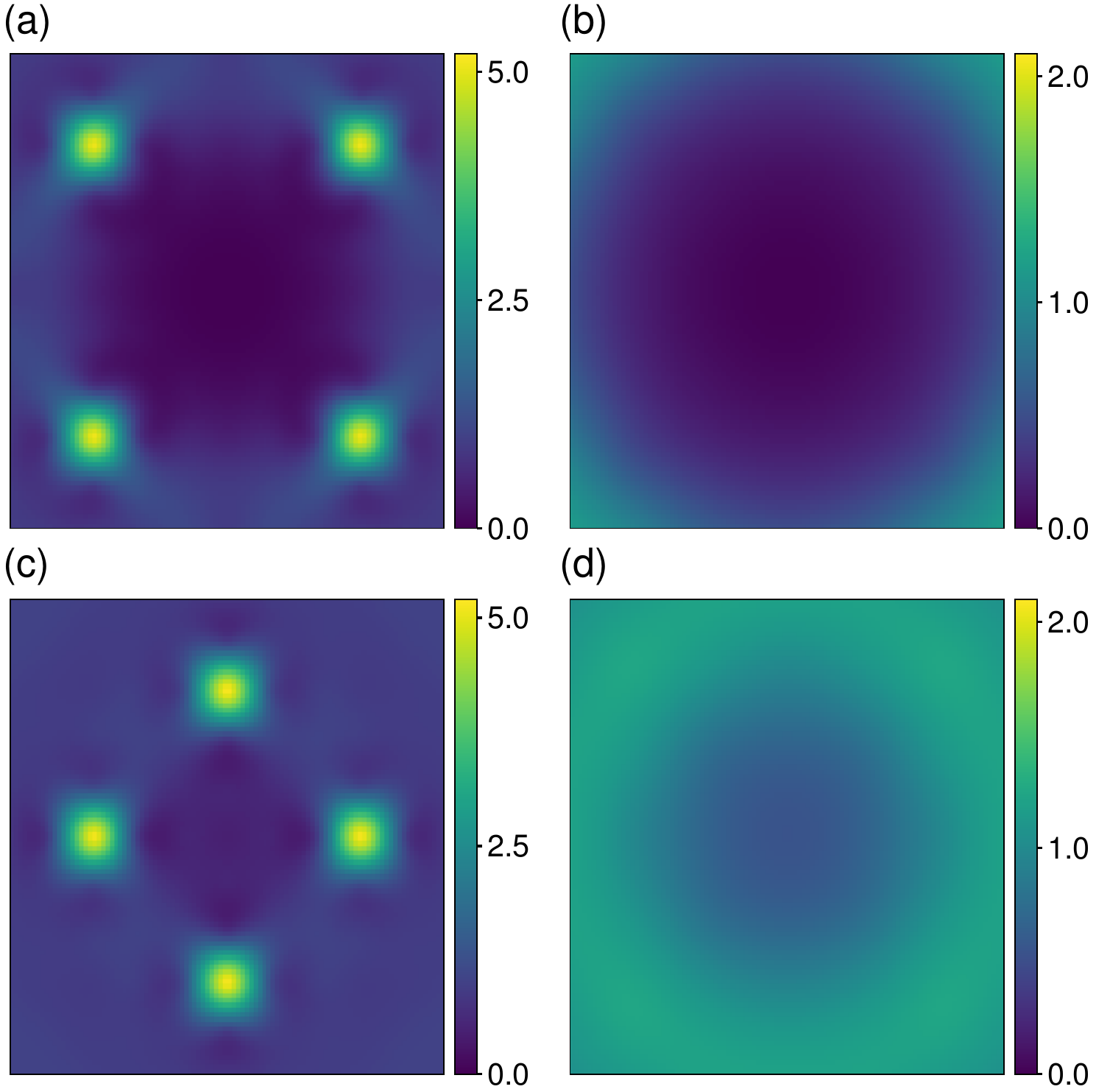}
\caption{Spatial distributions of the static structure factor \(S(\mathbf{k})\). Because the simulation cell employs periodic boundary conditions, the \(\mathbf{k}\) points are discrete; interpolation is therefore used to produce a continuous image of the distribution. Panels (a) and (b) show \(S(\mathbf{k})\) in the \(k_z = 0\) plane for \(r_s = 87\) and \(r_s = 50\), respectively. Panels (c) and (d) show \(S(\mathbf{k})\) in the \(k_z = k_{\mathrm{BCC}}/2\) plane for \(r_s = 87\) and \(r_s = 50\), respectively, where \(k_{\mathrm{BCC}}/2\) is half of the reciprocal lattice constant of the BCC unit cell.}
\label{fig:structure_factor}
\end{figure}

Here we present detailed correlation properties of the three-dimensional homogeneous electron gas in both the Fermi liquid (FL) and Wigner crystal (WC) phases. We use the MPNN wave function with the basis transformation for the unpolarized \(N=36\) electron system in a rectangular simulation cell with aspect ratio \(3:3:2\).

In Fig.~\ref{fig:real_correlation}, the spatial distributions of \(g(\mathbf{r})\) and \(g_s(\mathbf{r})\) are displayed. Panels (a) and (b) show \(g(\mathbf{r})\) and \(g_s(\mathbf{r})\) in the \(z=0\) plane for \(r_s=87\). Panels (c) and (d) show the same for \(r_s=50\) in the \(z=0\) plane. Panels (e) and (f) show \(g(\mathbf{r})\) and \(g_s(\mathbf{r})\) in the \(z=a_{\mathrm{BCC}}/2\) plane for \(r_s=87\), and panels (g) and (h) for \(r_s=50\). For the \(r_s=87\) case, the correlation functions distribution indicate that the electrons are in the WC phase and exhibit antiferromagnetic order on a BCC lattice. For \(r_s=50\), the correlation functions distribution show the characteristics of the FL phase.

In Fig.~\ref{fig:structure_factor}, we show the spatial distribution of \(S(\mathbf{k})\). Because the simulation cell uses periodic boundary conditions, the \(\mathbf{k}\) points are discrete; interpolation is used to produce a continuous image of the distribution. Panels (a) and (b) present the static structure factor \(S(\mathbf{k})\) in the \(k_z = 0\) plane for \(r_s = 87\) and \(r_s = 50\), respectively. Panels (c) and (d) show \(S(\mathbf{k})\) in the \(k_z = k_{\mathrm{BCC}}/2\) plane for \(r_s = 87\) and \(r_s = 50\), respectively, where \(k_{\mathrm{BCC}}\) is the reciprocal lattice constant of the BCC unit cell. At \(r_s = 87\), the plots reveal the characteristic Bragg peaks of the BCC crystal. At \(r_s = 50\), the system is in the FL phase and does not exhibit the crystalline Bragg-peak structure observed in the WC phase.

\subsection{Total Energy and Static Structure Factor Results}
In Tables~\ref{tab:supp_energy_table_fn} and~\ref{tab:supp_energy_table_mpnn}, we present the energies before and after the basis transformation for two network architectures at different \(r_s\) values, together with the optimized \(\alpha\) obtained from Step II. Table~\ref{tab:supp_energy_table_fn} corresponds to the FermiNet architecture for the unpolarized \(N=14\) electron system in a cubic cell. Table~\ref{tab:supp_energy_table_mpnn} corresponds to the MPNN architecture (with plane-wave reference state) for the unpolarized \(N=36\) electron system in a rectangular cell of aspect ratio \(3:3:2\). The uncertainty in \(\alpha\) is estimated as half the difference between the maximum and minimum values of \(\alpha\) during the stable oscillation regime after convergence: \(\delta \alpha \approx \frac{1}{2}(\alpha_{\text{max}} - \alpha_{\text{min}})\).

Table~\ref{tab:supp_energy_table_mpnn_scan} shows, for the MPNN architecture with the unpolarized \(N=36\) electron system in the same rectangular cell (\(3:3:2\)), the total energy \(E\) and the Bragg peak value of \(S(|\mathbf{k}|)\) near the phase transition, scanned over \(r_s\) for two reference states (PW and GO). Because the range of \(r_s\) is narrow, we fix \(\alpha\) for each reference state across different \(r_s\) values: \(\alpha_{\text{PW}} = 0.103\) for the PW reference state and \(\alpha_{\text{GO}} = 0.206\) for the GO reference state. These fixed \(\alpha\) values are those obtained from Step II optimization at \(r_s = 87\) for the respective reference states.

\begin{table}[ht]
\caption{
Total energies, in Hartree, for the unpolarized \(N=14\) electron system using the FermiNet architecture. Results are shown for \(n_{\text{det}} = 1, 4, 16\). \(E'\) denotes the energy obtained after step I, before applying the basis transformation; \(E\) denotes the energy obtained after step II. The dimensionless quantity \(r_s\sqrt{\alpha}\) is also shown, where \(\alpha\) is the optimized basis parameter.
}
\label{tab:supp_energy_table_fn}
\renewcommand{\arraystretch}{1.3}
\begin{tabular}{r | l l l | l l l| l l l}
\hline\hline
$r_s$ & $E'_{n_\text{det} = 1}$ & $E_{n_\text{det} = 1}$& $r_s\sqrt\alpha$ & $E'_{n_\text{det} = 4}$ & $E_{n_\text{det} = 4}$& $r_s\sqrt{\alpha}$ & $E'_{n_\text{det} = 16}$ & $E_{n_\text{det} = 16}$ & $r_s\sqrt{\alpha}$ \\[0.3ex]
\hline
1 & 7.96621(2) & 7.96622(6) & 25.4(4) & 7.96571(5) & 7.96570(7)& 25.6(3) & 7.96391(2) & 7.96391(2)& 33.4(9) \\[0.3ex]
2 & -0.116131(6) & -0.11614(1) & 21.6(4) & -0.116588(6) & -0.11659(2)& 25.3(5) & -0.117827(5) & -0.11783(1)& 26.1(5) \\[0.3ex]
5 & -1.117393(9) & -1.117469(6) & 22(1) & -1.117466(4) & -1.117537(5)& 22(2) & -1.118043(2) & -1.118082(3)& 23(2)  \\[0.3ex]
10 & -0.770977(3) & -0.771252(3) & 12.4(5) & -0.771216(4) & -0.771452(4)& 13.3(6) & -0.772378(3) & -0.772519(3)& 15.5(7)\\[0.3ex]
15 & -0.57083(5) & -0.57125(2) & 11.6(6) & -0.571157(8) & -0.571284(9)& 12.0(3) & -0.572009(2) & -0.572067(2)& 14.1(5)  \\[0.3ex]
20 & -0.45206(2) & -0.45224(1) & 10.2(5) & -0.452236(8) & -0.452367(3)& 10.2(5) & -0.45297(1) & -0.45304(4)& 11.7(6)\\
\hline\hline
\end{tabular}
\end{table}

\begin{table}[ht]
\caption{Total energy in Hartree for the unpolarized $N=36$ electron system using the MPNN architecture, with iteration numbers 1 and 2. \(E'\) denotes the energy obtained after step I, before applying the basis transformation; \(E\) denotes the energy obtained after step II. The dimensionless quantity \(r_s\sqrt{\alpha}\) is also shown, where \(\alpha\) is the optimized basis parameter.}
\label{tab:supp_energy_table_mpnn}
\renewcommand{\arraystretch}{1.3}
\begin{tabular}{r | l l l | l l l}
\hline\hline
$r_s$ & $E'_{\text{iteration} = 1}$ & $E_{\text{iteration} = 1}$& $r_s\sqrt\alpha$ & $E'_{\text{iteration} = 2}$ & $E_{\text{iteration} = 2}$& $r_s\sqrt{\alpha}$ \\[0.3ex]
\hline
5   & -2.72139(3)   & -2.72155(4)   & 19.3(8) & -2.72254(1)    & -2.72256(2)  & 23(2)  \\[0.3ex]
10  & -1.933647(8)  & -1.933741(9)  & 15.8(6) & -1.934601(2)   & -1.934627(4) & 22(1)  \\[0.3ex]
30  & -0.817604(2)  & -0.817621(3)  & 18(1)   & -0.8181171(4)  & -0.8181196(7)& 27(2)  \\[0.3ex]
50  & -0.5216283(8) & -0.521643(1)  & 18.2(4) &  -0.5219233(2) & -0.5219261(8)& 25(1)  \\[0.3ex]
80  & -0.3401625(3) & -0.3401708(9) & 18.0(2) & -0.3404184(6)  & -0.3404231(6)& 28.6(5)\\[0.3ex]
100 & -0.2767561(3) & -0.2767603(6) & 20.7(6) & -0.2769409(2)  & -0.2769434(4)& 24.5(2)\\
\hline\hline
\end{tabular}
\end{table}

\begin{table}[ht]
\caption{
Total energies, in Hartree, and Bragg-peak values for the unpolarized \(N=36\) electron system using the MPNN architecture with two message-passing iterations. Results are shown for PW and GO reference states. The basis parameter is fixed across the scanned \(r_s\) values for each reference state, with \(\alpha_{\mathrm{PW}}=0.103\) and \(\alpha_{\mathrm{GO}}=0.206\). Here \(E'\) and \(E\) denote the energies before and after the basis transformation, respectively, and \(S\) denotes the Bragg-peak value of the static structure factor after the basis transformation.
}
\label{tab:supp_energy_table_mpnn_scan}
\renewcommand{\arraystretch}{1.3}
\begin{tabular}{r | l l l | l l l}
\hline\hline
$r_s$ & $E'_{\text{PW}}$ & $E_{\text{PW}}$& $S_{\text{PW}}$ & $E'_{\text{GO}}$ & $E_{\text{GO}}$& $S_{\text{GO}}$ \\[0.3ex]
\hline
86.6 & -0.3164466(5) & -0.3164493(7) & 1.38(2) & -0.3164428(2) & -0.3164435(3)& 5.11(3) \\[0.3ex]
86.7 & -0.3161113(4) & -0.3161131(5) & 1.38(2) & -0.3161083(2) & -0.3161087(3) & 5.11(3)  \\[0.3ex]
86.8 & -0.3157748(4) & -0.3157773(5) & 1.39(2) & -0.3157743(2) & -0.3157745(3)& 5.15(3)  \\[0.3ex]
86.9 & -0.3154400(3) & -0.3154422(5) & 1.39(2) & -0.3154413(2) & -0.3154419(3)& 5.12(3)\\[0.3ex]
87.0 & -0.3151011(4) & -0.3151048(4) & 1.38(2) & -0.3151082(2) & -0.3151086(3)& 5.13(3)\\[0.3ex]
87.1 & -0.3147681(5) & -0.3147713(7) & 1.46(2) & -0.3147772(2) & -0.3147776(3)& 5.14(3)\\
\hline\hline
\end{tabular}
\end{table}

\clearpage
\newpage